\documentclass[11pt]{article}

\usepackage{graphicx}
  \DeclareGraphicsExtensions{.jpeg,.png,.jpg,.eps}
\usepackage{epstopdf} 
\usepackage{amsmath,amssymb,amsfonts,epsfig,hyperref,color}

\newtheorem{theorem}{Theorem}[section]
\newtheorem{corollary}{Corollary}[theorem]

\def\eb{{\bf e}}

\def\R{{\mathbb R}}

\def\D{\Delta}

\def\l{\lambda}

\def\ap{\rightarrow}

\def\sg{\mbox{sign}}
\def\nm{\Vert}

\newcommand{\bc}{\begin{center}}
\newcommand{\ec}{\end{center}}
\newcommand{\be}{\begin{equation}}
\newcommand{\ee}{\end{equation}}
\newcommand{\bd}{\begin{displaymath}}
\newcommand{\ed}{\end{displaymath}}
\newcommand{\ba}{\begin{array}}
\newcommand{\ea}{\end{array}}
\newcommand{\ben}{\begin{enumerate}}
\newcommand{\een}{\end{enumerate}}
\newcommand{\bit}{\begin{itemize}}
\newcommand{\eit}{\end{itemize}}
\newcommand{\beq}{\begin{eqnarray}}
\newcommand{\eeq}{\end{eqnarray}}
\newcommand{\btab}{\begin{tabular}}
\newcommand{\etab}{\end{tabular}}
\newcommand{\bfig}{\begin{figure}}
\newcommand{\efig}{\end{figure}}
\newcommand{\btp}{\begin{tikzpicture}}
\newcommand{\etp}{\end{tikzpicture}}

\newcommand{\nmm}[1]{ \nm #1 \nm }
\newcommand{\nmeu}[1]{ \nm #1 \nm_2 }

\def\nmsl1{\nm_{{\rm SL1}}}

\begin{document}


\title{CLOT Norm Minimization for
Continuous Hands-off Control
} 



\author{Niharika Challapalli, Masaaki Nagahara and
Mathukumalli Vidyasagar
\thanks{NC is with the
Department of Electrical Engineering,
University of Texas at Dallas,
Richardson, TX 75080, USA
(e-mail: niharika15c@gmail.com).
MN is with the
Institute of Environmental Science and Technology,
The University of Kitakyushu,
Hibikino 1-1, Wakamatsu-ku, Kitakyushu, Fukuoka
808-0135, JAPAN (e-mail: nagahara@kitakyu-u.ac.jp, nagahara@ieee.org).
MV is with the
Department of Systems Engineering,
University of Texas at Dallas,
Richardson, TX 75080, USA, and
Department of Electrical Engineering,
Indian Institute of Technology Hyderabad,
Kandi, Telangana, India 502285 
(e-mail: m.vidyasagar@utdallas.edu, m.vidyasagar@iith.ac.in).
The research of MN was supported in part by JSPS
KAKENHI Grant Numbers 15H02668, 15K14006, and 16H01546.
The research of MV and NC was supported
by the US National Science Foundation under Award No.\
ECCS-1306630, the Cancer Prevention and Research
Institute of Texas (CPRIT) under award No.\ RP140517, and
a grant from the Department of Science and Technology,
Government of India.
}
}

\maketitle

\begin{abstract}                
In this paper, we consider hands-off control via minimization of the CLOT
(Combined $L$-One and Two) norm.
The maximum hands-off control is the $L^0$-optimal (or the sparsest) control among all feasible controls
that are bounded by a specified value and transfer the state from a given initial state to the origin within a fixed time duration.
In general, the maximum hands-off control is a bang-off-bang control taking values of $\pm 1$ and $0$.
For many real applications, such discontinuity in the control is not desirable.
To obtain a continuous but still relatively
sparse control, we propose to use the CLOT norm,
a convex combination of $L^1$ and $L^2$ norms.
We show by numerical simulations that the CLOT control is continuous and much sparser
(i.e. has longer time duration on which the control takes 0) than the conventional EN (elastic net) control,
which is a convex combination of $L^1$ and squared $L^2$ norms.
We also prove that the CLOT control is continuous in the sense that, if
$O(h)$ denotes the sampling period, then the difference between
successive values of the CLOT-optimal control is 
$O(\sqrt{h})$, which
is a form of continuity. Also, the CLOT formulation is extended
to encompass constraints on the state variable.
\end{abstract}



\section{Introduction}
\label{sec:Introduction}

Sparsity has recently emerged as an important topic in signal/image processing, machine learning, statistics, etc.
If $y \in \R^m$ and $A \in \R^{m \times n}$ are specified with $m < n$, then
the equation $y = Ax$ is underdetermined and has infinitely many solutions
for $x$ if $A$ has rank $m$.
Finding the sparsest solution (that is, the solution with the fewest
number of nonzero elements) can be formulated as
\[
 \min_{z} \|z\|_0 \mathrm{~~subject~to~~} Az=b.
\]
However, this problem is NP hard, as shown in \cite{Nat95}.
Therefore other approaches have been proposed for this purpose.
This area of research is known as ``sparse regression.''
One of the most popular is LASSO \cite{Tib96}, also referred to as
forgetting \cite{Ish96}, or basis pursuit \cite{CheDonSau99}, in which the $\ell^0$-norm is
replaced by the $\ell^1$-norm.
Thus the problem becomes
\[
 \min_{z} \|z\|_1 \mathrm{~~subject~to~~} Az=b.
\]
The advantage of LASSO is that it is a convex optimization problem and
therefore very large problems can be solved efficiently, for example
by using the Matlab-based package \texttt{cvx} \cite{GraBoy14}.
Moreover, under mild technical assumptions, the LASSO-optimal solution
has no more than $m$ nonzero components \cite{Osborne-Presnell-Turlach00}.
However, the exact location of the nonzero components is very sensitive
to the vector $y$.
To overcome this deficiency, another approach known as the Elastic Net
was proposed in \cite{ZouHas05}, where the $\ell^1$ norm in LASSO
is replaced by
a weighted sum of $\ell^1$ and squared $\ell^2$ norms.
This leads to the optimization problem
\[
 \min_{z} \lambda_1\|z\|_1+\lambda_2\|z\|_2^2 \mathrm{~~subject~to~~} Az=b,
\]
where $\lambda_1$ and $\lambda_2$ are positive weights such that $\lambda_1+\lambda_2=1$.
It is shown in \cite[Theorem 1]{ZouHas05}
that the EN formulation gives the \emph{grouping effect};
If two columns of the matrix $A$ are highly correlated, then the
corresponding components of the solution for $x$ have nearly equal values.
This ensures that the solution for $x$ is not overly sensitive to small
changes in $y$.
The name ``elastic net'' is meant to suggest
a stretchable fishing net that retains \emph{all the big fish}.

During the past decade and a half, another research area known as
``compressed sensing'' has witnessed a great deal of interest.
In compressed sensing, the matrix $A$ is not specified; rather, the user
gets to choose the integer $m$ (known as the number of measurements),
as well as the matrix $A$.
The objective is to choose the matrix $A$ as well as a corresponding
``decooder'' map $\D: \R^m \ap \R^n$ such that, 
the unknown vector $x$ is sparse and the measurement vector $y$
equals $Ax$, then $\D(Ax) = x$ for all sufficiently sparse vectors $x$.
More generally,
if measurement vector $y = Ax + \eta$ where $\eta$ is the
measurement noise, and the vector $x$ is nearly sparse (but not exactly
sparse),
then the recovered vector $\D(Ax + \eta)$ should be sufficiently
close to the true but unknown vector $x$.
This is referred to as ``robust sparse recovery.''
Minimizing the $\ell_1$-norm is among the more popular decoders.
See the books by \cite{Ela}, \cite{EldKut}, and \cite{FouRau} for the
theory and some applications.
Due to its similarity to the LASSO formulation of \cite{Tib96},
this approach to compressed sensing is also referred to as LASSO.

Until recently the situation was that LASSO achieves robust sparse recovery
in compressed sensing, but did not achieve the grouping effect in
sparse regression.
On the flip side, EN achieves the grouping effect, but it was not known
whether it achieves robust sparse recovery.
A recent paper \cite{AhsChaVid16} sheds some light on this problem.
It is shown in \cite{AhsChaVid16} that EN \textit{does not achieve}
robust sparse recovery.
To achieve both the grouping effect in sparse regression as well as
robust sparse recovery in compressed sensing,
\cite{AhsChaVid16} has proposed the CLOT (Combined $L$-One and Two) formulation:
\[
 \min_{z} \lambda_1\|z\|_1+\lambda_2\|z\|_2 \mathrm{~~subject~to~~} Az=b,
\]
where $\l_1>0$, $\l_2>0$, and $\l_1 + \l_2 = 1$.
The difference between EN and CLOT is the $\ell^2$ norm term;
EN has the squared $\ell^2$ norm while CLOT has the pure $\ell^2$ norm.
This slight change leads to both the grouping effect and robust sparse recovery,
as shown in \cite{AhsChaVid16}.

In parallel with these advances in sparse regression and recovery of
unknown sparse vectors, 
sparsity techniques have also been applied to control.
Sparsity-promoting optimization has been applied to networked
control in \cite{NagQueOst14}, where
quantization errors and data rate can be reduced at the same time
by sparse representation of control packets.
Other examples of control applications include
optimal controller placement by \cite{CasClaKun12,ClaKun12,FarLinJov11},
design of feedback gains by \cite{FuFarJov13,PolKhlShc13},
state estimation by \cite{ChaAsiRomRoz11},
to name a few. 

More recently, 
a novel control called the {\em maximum hands-off control} has been 
proposed in~\cite{NagQueNes16} for \emph{continuous-time} systems.
The maximum hands-off control 
is the $L^0$-optimal control (the control that has the minimum support length) among all feasible controls
that are bounded by a fixed value and transfer the state from a given initial state to the origin within a fixed time duration.
Such a control is effective for
reduction of electricity or fuel consumption; an electric/hybrid vehicle
shuts off the internal combustion engine (i.e. hands-off control)
when the vehicle is stopped or the speed is lower than a preset threshold; see \cite{Cha07} for example.
Railway vehicles also utilize hands-off control, often called {\em coasting control},
to cut electricity consumption; see \cite{LiuGol03} for details.
In~\cite{NagQueNes16}, the authors have proved the theoretical relation between 
the maximum hands-off control and the $L^1$ optimal control 
under the assumption of normality.
Also, important properties of the maximum hands-off
control have been proved
in \cite{IkeNag16_automatica} for
the convexity of the value function,
and in \cite{ChaNagQueRao16} 
for necessary conditions of optimality,
and in \cite{IkeNagOno17} for the discreteness.

In general, the maximum hands-off control is a bang-off-bang control taking values of $\pm 1$ and $0$.
For many real applications, such a discontinuity property is not desirable.
To obtain a continuous but still sparse control, 
\cite{NagQueNes16} has proposed to use a combined $L^1$ and \emph{squared} $L^2$ minimization,
like EN mentioned above.
Let us call this control an EN control.
As in the case of EN in the vector optimization, the EN control often shows much less sparse
(i.e. has a larger $L^0$ norm) than the maximum hands-off control.
Then, in \cite{CLOT_Control:IFAC17}, we have proposed to use the CLOT norm,
a convex combination of $L^1$ and \emph{non-squared} $L^2$ norms.
The minimum CLOT-norm control is called the CLOT control.
In \cite{CLOT_Control:IFAC17}, we have shown by numerical simulation that the CLOT control is continuous and much sparser
(i.e. has longer time duration on which the control takes 0) than the conventional EN control.

In \cite{NagQueNes16}, both the LASSO and EN approaches to hands-off
control are solved in continuous-time.
It is shown, using Pontryagin's minimum principle, that the LASSO
solution is bang-off-bang, while the EN solution is continuous.
However, the CLOT formulation cannot be addressed via Pontryagin's principle.
Therefore it is not clear whether the resulting optimal control is
continuous.
In the present paper, we study the discretized problem, and show that
as the sampling interval $h$ approaches zero, the difference between
successive control signals is $O(\sqrt{h})$, which is a form of
continuity.
We extend this result to the case where, in addition to constraints on
the control signal, there are also constraints on the state $x(t)$.

The remainder of this article is organized as follows.
In Section \ref{sec:Problem Formulation}, we formulate the control problem considered in this paper.
In Section \ref{sec:Discretization}, we give a discretization method to numerically compute the optimal control.
In Section \ref{sec:state_constraints}, we give the additional state constraints for the optimization problems.
The limiting behaviour of the CLOT optimal control is stated in Section \ref{sec: conti_CLOT}. 
Results of the numerical computations on a variety of problems without any state constriants are presented in Section \ref{sec:exam}.
Results of the numerical computations on a variety of problems with state constriants are presented in Section \ref{sec:exam}.
These examples illustrate the advantages of the CLOT control
compared with the maximum hands-off control and the EN control.
We present some conclusions in Section \ref{sec:Conclusions}.
\subsection*{Notation}

Let $T>0$ and $m\in\mathbb{N}$. 
For a continuous-time signal $u(t)\in\mathbb{R}$ over a time interval $[0, T]$, 
we define its $L^p$ ($p\geq 1$) and $L^{\infty}$ norms respectively by
\[
 \|u\|_{p} \triangleq \bigg\{\int_{0}^{T} |u(t)|^{p} dt\bigg\}^{1/p},~
 \|u\|_{\infty} \triangleq \sup_{t\in[0, T]}|u(t)|.
\]
We denote the set of all signals with $\|u\|_{p}<\infty$ by $L^p[0,\,T]$
for $p\geq 1$ or $p=\infty$.
We define the $L^0$ norm of a signal $u(t)$ on the interval $[0, T]$ as
\[
  \|u\|_{0}\triangleq \int_{0}^{T}\phi_{0}(u(t)) dt,
\]
where $\phi_{0}$ is the $L^0$ kernel function defined by
\begin{equation}
  \phi_0(\alpha) \triangleq 
    \begin{cases} 
    1, & \text{~if~} \alpha\neq 0,\\ 
    0, & \text{~if~} \alpha=0
    \end{cases}
\label{eq:L0-kernel}
\end{equation}
for a scalar $\alpha\in\mathbb{R}$.
The $L^0$ norm can be represented by
\[
 \|u\|_0 = \mu_{\mathrm{L}}\bigl(\mathrm{supp}(u)\bigr),
\]
where $\mathrm{supp}(u)$ is the support of the signal $u$, and
$\mu_{\mathrm{L}}$ is the Lebesgue measure on ${\mathbb{R}}$.
\section{Problem Formulation}
\label{sec:Problem Formulation}

Let us consider a linear time-invariant system described by
\begin{equation}
 \frac{dx}{dt}(t) = Ax(t) + Bu(t),~ t\geq 0,~ x(0)=\xi.
 \label{eq:plant}
\end{equation}
Here we assume that $x(t)\in{\mathbb{R}}^n$, $u(t)\in{\mathbb{R}}$,
and the initial state $x(0)=\xi$ is fixed and given.
The control objective is to drive the state $x(t)$ from $x(0)=\xi$ to the origin
at time $T>0$, that is
\begin{equation}
 x(T)=0.
 \label{eq:terminal condition}
\end{equation}
We limit the control $u(t)$ to satisfy
\begin{equation}
 \|u\|_\infty \leq U_{\max}
 \label{eq:u_max}
\end{equation}
for fixed $U_{\max}>0$.

If the system \eqref{eq:plant} is controllable and the final time $T$ is larger than
the optimal time $T^\ast$ (the minimal time in which there exist a control $u(t)$
that drives $x(t)$ from $x(0)=\xi$ to the origin; see \cite{HerLas}), 
then there exists at least one
$u(t)\in L^\infty[0,T]$ that satisfies
equations \eqref{eq:plant}, \eqref{eq:terminal condition}, and \eqref{eq:u_max}.
Let us call such a control a \emph{feasible} control.
From \eqref{eq:plant} and \eqref{eq:terminal condition},
any feasible control $u(t)$ on $[0,T]$ satisfies
\[
 0=x(T)=e^{AT}\xi + \int_0^T e^{A(T-t)}Bu(t)dt,
\]
or
\begin{equation}
 \int_0^T e^{-At}Bu(t)dt+\xi=0.
 \label{eq:feasible control}
\end{equation}
Define a linear operator $\Phi:L^\infty[0,T] \rightarrow {\mathbb{R}}^n$ by
\[
 \Phi u \triangleq \int_0^T e^{-At}Bu(t)dt,\quad u\in L^\infty[0,T].
\]
By this, we define the set ${\mathcal{U}}$ of the feasible controls by
\begin{equation}
 \mathcal{U}\triangleq
  \left\{u\in L^\infty: \Phi u+ \xi=0,~\|u\|_\infty\leq 1\right\}.
  \label{eq:feasible controls}
\end{equation}

The problem of the maximum hands-off control
is then described by
\begin{equation}
 \underset{u}{\mathrm{minimize}}~~ \|u\|_0 \mathrm{~~subject~to~~} u \in \mathcal{U}.
 \label{eq:L0 optimal control}
\end{equation}

The $L^0$ problem \eqref{eq:L0 optimal control} is very hard to solve since the $L^0$ cost function
is non-convex and discontinuous.
For this problem, \cite{NagQueNes16} has shown that the $L^0$ optimal control in \eqref{eq:L0 optimal control} is
equivalent to the following $L^1$ optimal control:
\begin{equation}
 \underset{u}{\mathrm{minimize}}~~ \|u\|_1 \mathrm{~~subject~to~~} u \in \mathcal{U},
 \label{eq:LASSO}
\end{equation}
\emph{if} the plant is normal, that is, if the system \eqref{eq:plant} is 
controllable and the matrix $A$ is nonsingular.
Let us call the $L^1$ optimal control as the \emph{LASSO control}.
If the plant is normal, then the LASSO control is in general a \emph{bang-off-bang} control
that is piecewise constant taking values in $\{0,\pm 1\}$.
The discontinuity of the LASSO solution is not desirable in real applications,
and a smoothed solution is also proposed in \cite{NagQueNes16} as
\begin{equation}
 \underset{u}{\mathrm{minimize}}~~ \|u\|_1+\lambda\|u\|_2^2 \mathrm{~~subject~to~~} u \in \mathcal{U},
 \label{eq:EN}
\end{equation}
where $\lambda>0$ is a design parameter for smoothness.
Let us call this control the \emph{EN (elastic net) control}.
In \cite{NagQueNes16}, it is proved that the solution of \eqref{eq:EN} is
a continuous function on $[0,T]$.

While the EN control is continuous, it is shown by numerical experiments that
the EN control is not sometimes sparse.
This is an analogy of the EN for finite-dimensional vectors that
EN does not achieve robust sparse recovery.
Borrowing the idea of CLOT in \cite{AhsChaVid16},
we define the CLOT optimal control problem by
\begin{equation}
 \underset{u}{\mathrm{minimize}}~~ \|u\|_1+\lambda\|u\|_2 \mathrm{~~subject~to~~} u \in \mathcal{U}.
 \label{eq:CLOT}
\end{equation}
We call this optimal control the \emph{CLOT control}.
\section{Discretization}
\label{sec:Discretization}

Since the problems \eqref{eq:LASSO}--\eqref{eq:CLOT} are infinite dimensional,
we should approximate it to finite dimensional problems.
For this, we adopt the time discretization.

First, we divide the time interval $[0,T]$ into $N$ subintervals,
$[0,T] = [0,h) \cup \dots \cup [(N-1)h,Nh]$,
where $h$ is the discretization step (or the sampling period)
such that $T=Nh$.
We assume that the state $x(t)$ and the control $u(t)$ in \eqref{eq:plant}
are constant over each subinterval.
On the discretization grid,
$t=0,h,\dots,Nh$,
the continuous-time system \eqref{eq:plant} is described as
\begin{equation} \label{eq:dis_states}
 \hat{x}_{k+1} = A_d \hat{x}_k + B_d \hat{u}_k,\quad k=0,1,\dots,N-1,
\end{equation}
where $\hat{x}_k\triangleq x(kh)$, $\hat{u}_k\triangleq u(kh)$, and
\begin{equation}\label{eq:AdBd}
 A_d \triangleq e^{Ah},\quad B_d \triangleq \int_0^h e^{At}Bdt.
\end{equation}
Define the control vector
\begin{equation}
 \hat{u} \triangleq [\hat{u}_0, \hat{u}_1,\dots,\hat{u}_{N-1}]^\top.
\end{equation} 
Note that the final state $x(T)$ can be described as
\begin{equation}
 x(T)=\hat{x}_N=A_d^N \xi + \Phi_N \hat{u},
\end{equation}
where
\begin{equation}
 \Phi_N \triangleq \begin{bmatrix}A_d^{N-1}B_d,&A_d^{N-2}B_d,&\dots,&B_d\end{bmatrix}.
 \label{eq:Phi_N}
\end{equation}
Then the set $\mathcal{U}$ in \eqref{eq:feasible controls} is approximately
represented by
\begin{equation}
 {\mathcal{U}}_N \triangleq
 \left\{\hat{u}\in {\mathbb{R}}^N: A_d^N\xi + \Phi_N \hat{u} =0,~\|\hat{u}\|_\infty\leq 1\right\}.
 \label{eq:approximated feasible controls}
\end{equation}

Next, we approximate the $L^1$ norm of $u$ by
\begin{equation}
 \begin{split}
 \|u\|_1 &= \int_0^T |u(t)| dt\\
 &= \sum_{k=0}^{N-1} \int_{kh}^{(k+1)h} |u(t)| dt\\
 &\approx \sum_{k=0}^{N-1} \int_{kh}^{(k+1)h} |\hat{u}_k| dt\\
 &= \sum_{k=0}^{N-1} |\hat{u}_k| h\\
 &= \|\hat{u}\|_1 h.
 \end{split}
\end{equation}
In the same way, we obtain approximation of the $L^2$ norm of $u$ as
\begin{equation}
\|u\|_2^2 = \int_0^T |u(t)|^2 dt\approx \|\hat{u}\|_2^2 h.
\end{equation}

Finally, the optimal control problems \eqref{eq:LASSO},
\eqref{eq:EN} and \eqref{eq:CLOT} can be approximated by
\begin{align}
 &\underset{\hat{u}\in{\mathbb{R}}^N}{\mathrm{minimize}}~~
  h\|\hat{u}\|_1 \mathrm{~~subject~to~~} \hat{u}\in{\mathcal{U}}_N,
 \label{eq:LASSO_d}\\
&\underset{\hat{u}\in{\mathbb{R}}^N}{\mathrm{minimize}}~~
h\|\hat{u}\|_1+h\lambda\|\hat{u}\|_2^2
\mathrm{~~subject~to~~}
\hat{u}\in{\mathcal{U}}_N,
 \label{eq:EN_d}\\
& \underset{\hat{u}\in{\mathbb{R}}^N}{\mathrm{minimize}}~~
h\|\hat{u}\|_1+\sqrt{h}\lambda \|\hat{u}\|_2
\mathrm{~~subject~to~~}
\hat{u}\in{\mathcal{U}}_N. \label{eq:CLOT_d}
\end{align}
The optimization problems
are convex and can be efficiently solved by
numerical software packages such as \verb=cvx= with Matlab;
see \cite{GraBoy14} for details.
\section{Optimal control with additional state contraints}\label{sec:state_constraints}
In this section, additional constraints are introducted to optimization problems \eqref{eq:LASSO_d}, \eqref{eq:EN_d} and \eqref{eq:CLOT_d} on the states to ensure that $\ell_2$ norm of the state at any given instant does not blow up.

The constraint is the $\ell_2$ norm of the state vector at any given time should not exceed a specified threshold $\theta$, that is,
\begin{equation}
\nmm{\hat{x}_k }_2 \leq \theta, \qquad k \in\{1,2,\ldots,N-1\},
\end{equation}
where $\hat{x}_k$ is the discrete-time state at time instant $k$ as defined in \eqref{eq:dis_states}.
Using \eqref{eq:dis_states}, the states are described as
\begin{equation}
\begin{bmatrix}
\hat{x}_1 \\
\hat{x}_2 \\
\hat{x}_3 \\
\vdots \\
\hat{x}_{N-1}
\end{bmatrix} =
\begin{bmatrix}
A_d \\
A_d^2 \\
A_d^3 \\
\vdots \\
A_d^{N-1}
\end{bmatrix} \xi + \Psi_N 
\begin{bmatrix}
\hat{u}_0 \\
\hat{u}_1 \\
\hat{u}_2 \\
\vdots \\
\hat{u}_{N-2}
\end{bmatrix}
\end{equation}
where
\begin{equation}
\Psi_N =
\begin{bmatrix}
B_d & 0 & 0 & \dots & \dots & 0 \\
A_d B_d & B_d & 0 & \dots & \dots & 0 \\
A_d^2 B_d & A_d B_d & B_d & \dots & \dots & 0 \\
\dots \\
\dots \\
A_d^{N-2} B_d & A_d^{N-3} B_d & A_d^{N-4} B_d & \dots & \dots & B_d 
\end{bmatrix}
\end{equation}
and $\Psi_N \in \mathbb{R}^{(N-1)n \times (N-1)}$.
\subsection{How to choose $\theta$?}

The following steps are followed in order to choose $\theta$:
\begin{enumerate}
\item First, we solve the control optimization problems without state constraint and then we note the maximum of $\ell_2$ norm of the state vector, say $l_{\max}$.
\item It is to be noted that if $\theta \geq l_{\max}$, the problem is still unconstrained with respect to state. Thus, maximum value of the threshold ($\theta_{\max}$) is $l_{\max}$.
\item Then, we set $\theta$ to $\theta_{\max}$ and keep decreasing the value of $\theta$ until the optimization problems become infeasible. This gives us lower bound on $\theta$.
\item Since, the state constraints are dependent on the system, we get a range of $\theta$ that is specific to each problem.
\end{enumerate}

Therefore, the optimization problems respectively become
\begin{equation}\label{eq:LASSO_state}
\begin{aligned}
 & \underset{\hat{u}\in{\mathbb{R}}^N}{\mathrm{minimize}}
 & & h\|u\|_1 \\
 & \mathrm{subject~to}
 & & \hat{u}\in{\mathcal{U}}_N,~~ \nmm{\hat{x}_k }_2 \leq \theta\\
 &&& k \in \{1,\cdots,N-1\}
\end{aligned}  
\end{equation}
for LASSO control,
\begin{equation}\label{eq:EN_state}
\begin{aligned}
 & \underset{\hat{u}\in{\mathbb{R}}^N}{\mathrm{minimize}}
 & & h\|\hat{u}\|_1+h\lambda\|\hat{u}\|_2^2 \\
 & \mathrm{subject~to} 
 & & \hat{u}\in{\mathcal{U}}_N,~~ \nmm{\hat{x}_k }_2 \leq \theta\\
 &&& k \in \{1,\cdots,N-1\}
\end{aligned}  
\end{equation}
for EN control, and
\begin{equation}\label{eq:CLOT_state}
\begin{aligned}
 &\underset{\hat{u}\in{\mathbb{R}}^N}{\mathrm{minimize}}
 & & h\|\hat{u}\|_1+\sqrt{h}\lambda \|\hat{u}\|_2 \\
 & \mathrm{subject~to}
 & & \hat{u}\in{\mathcal{U}}_N,~~\nmm{\hat{x}_k }_2 \leq \theta\\
 &&& k \in \{1,\cdots,N-1\}
\end{aligned}
\end{equation}
for CLOT control. 

\section{Limiting behavior of CLOT solution}\label{sec: conti_CLOT}

In this section, we show the limiting behaviour of CLOT optimal control. 

\begin{theorem}\label{thm: CLOT_conti}
	If $\hat{u}$ is the solution of the problem \eqref{eq:CLOT_state}, 
	then $| \hat{u}_k - \hat{u}_{k+1} |$ is of $O( \sqrt{h} )$,
	where $\hat{u}_k$ is the $k$-th entry of $\hat{u}$.
\end{theorem}

Due to the length of the proof, it is added to the appendix \ref{app: proof_conti}. It can be noted that the slope of $\hat{u}$ is of $O({1}/\sqrt{h})$, thus as $h \rightarrow 0$, the slope blows up, thus the CLOT optimal control closely approximates $L^1$ optimal control. Therefore, CLOT optimal control solution is continuous approximation of $L^1$ optimal control.
\begin{corollary}
If $\hat{u}$ is the solution of the problem \eqref{eq:CLOT_state} 
without state constraints (i.e. $\theta$ is sufficiently large), 
then $| \hat{u}_k - \hat{u}_{k+1} |$ is of $O( \sqrt{h} )$.	
\end{corollary}

\section{Numerical Examples without state constraints}
\label{sec:exam}

In this section we present numerical results from applying the CLOT norm
minimization approach to seven different plants, and compare the results
with those from applying LASSO and EN.
\subsection{Details of Various Plants Studied}

For the reader's convenience, the details of the various plants are
given in Table \ref{table:plants}.
The figure numbers show where the corresponding
computational results can be found.
Some conventions are adopted to reduce the clutter in the table, as
described next.
All plants are of the form
\bd
P(s) = \frac{n(s)}{d(s)}, n(s) = \prod_{i=1}^{n_z} (s - z_i ) ,
d(s) = \prod_{i=1}^{n_p} (s - p_i) .
\ed
To save space in the table, the plant zeros are not shown;
$P_3(s)$ has a zero at $s = -2$, $P_6(s)$ has a zero at $s = 2$,
while $P_7(s)$ has zeros at $s = 1, 2$.
The remaining plants do not have any zeros, so that the plant
numerator equals one.

Once the plant zeros and poles are specified, the plant numerator
and denominator polynomials $n,d$ were computed using the Matlab command
{\tt poly}.
Then the transfer function was computed as {\tt P = tf(n,d)},
and the state space realization was computed as
{\tt [A,B,C,D] = ssdata(P)}.
The maximum
control amplitude is taken $1$, so that the control must satisfy
$|u(t)| \leq 1$ for $t \in [0,T]$.
To save space, we use the notation $\eb_l$ to denote an $l$-column
vector whose elements all equal one.
Note that in all but one case, the initial condition equals $\eb_n$
where $n$ is the order of the plant.

Note that, with $T = 20$, the problems with plants $P_6(s)$
and $P_7(s)$ are not feasible (meaning that $T$ is smaller than
the minimum time needed to reach the origin); this is why we took $T = 40$.

All optimization problems were solved after discretizing the interval
$[0,T]$ into both 2,000 as well as 4,000 samples, to examine whether the
sampling time affects the sparsity density of the computed optimal control.

\begin{table}
\bc
\btab{|l|l|l|l|l|l|l|}
\hline
\!\textbf{No}\!&\textbf{Plant}\!&\!\textbf{Poles}\!&\!$T$\!&\!$x(0)$\!&\!$\l$\!
&\!\textbf{Figs}\!\\
\hline
1 & $P_1(s)$ & 0,0,0,0 & 20 & $\eb_4$ & 1 & 
\ref{fig:P1_state_l_1}, \ref{fig:P1_control_l_1} \\
\hline
2 & $P_1(s)$ & 0,0,0,0 & 20 & $\eb_4$ & 0.1 &
\ref{fig:P1_state_l_01}, \ref{fig:P1_control_l_01} \\
\hline
3 & $P_2(s)$ & $ -0.025 \pm j$ & 20 & $\eb_2$ & 0.1 &
\ref{fig:P2_state_1_1}, \ref{fig:P2_control_1_1} \\
\hline
4 & $P_2(s)$ & $ -0.025 \pm j$ & 20 & $(10,1)^\top$ & 0.1 &
\ref{fig:P2_state_10_1}, \ref{fig:P2_control_10_1} \\
\hline
5 & $P_3(s)$ & $-1 \pm 0.2j$ & 20 & $\eb_4$ & 0.1 & 
\ref{fig:P3_state}, \ref{fig:P3_control} \\
 & &$\pm j$& & & & \\
\hline
6 & $P_4(s)$ & $-5 \pm j$ & 20 & $\eb_6$ & 0.1 & 
\ref{fig:P5_state}, \ref{fig:P5_control} \\
& & $-0.3 \pm 2j$ & & & & \\
& & $-1 \pm 2 \sqrt{2} j$ & & & & \\
\hline
7 & $P_5(s)$ & $0,0,0,0$ & 40 & $\eb_6$ & 0.1 & 
\ref{fig:P6_state}, \ref{fig:P6_control} \\
 & &$\pm j$& & & & \\
\hline
8 & $P_6(s)$ & $0,0,0,0$ & 40 & $\eb_6$ & 0.1 & 
\ref{fig:P7_state}, \ref{fig:P7_control} \\
 & &$\pm j$& & & & \\
\hline
\etab
\ec
\caption{Details of various plants studied}
\label{table:plants}
\end{table}
\subsection{Plots of Optimal State and Control Trajectories}

The plots of the $\ell^2$-norm (or Euclidean norm)
of the state vector trajectory and the
control signal for all these examples are shown in the next several plots.

We begin with the plant $P_1(s)$, the fourth-order integrator.
Figures \ref{fig:P1_state_l_1} and \ref{fig:P1_control_l_1}
show the state and control trajectories when $\l = 1$.
The same system is analyzed using a smaller value of $\l = 0.1$.
One would expect that the resulting control signals would be more sparse
with a smaller $\l$, and this is indeed the case.
The results are shown in 
Figures \ref{fig:P1_state_l_01} and \ref{fig:P1_control_l_01}.
Based on the observation that the control signal becomes more sparse
with $\l = 0.1$ than with $\l = 1$, all the other plants are analyzed
with $\l = 0.1$.
	
\bfig
		\bc
		\includegraphics[width=80mm]{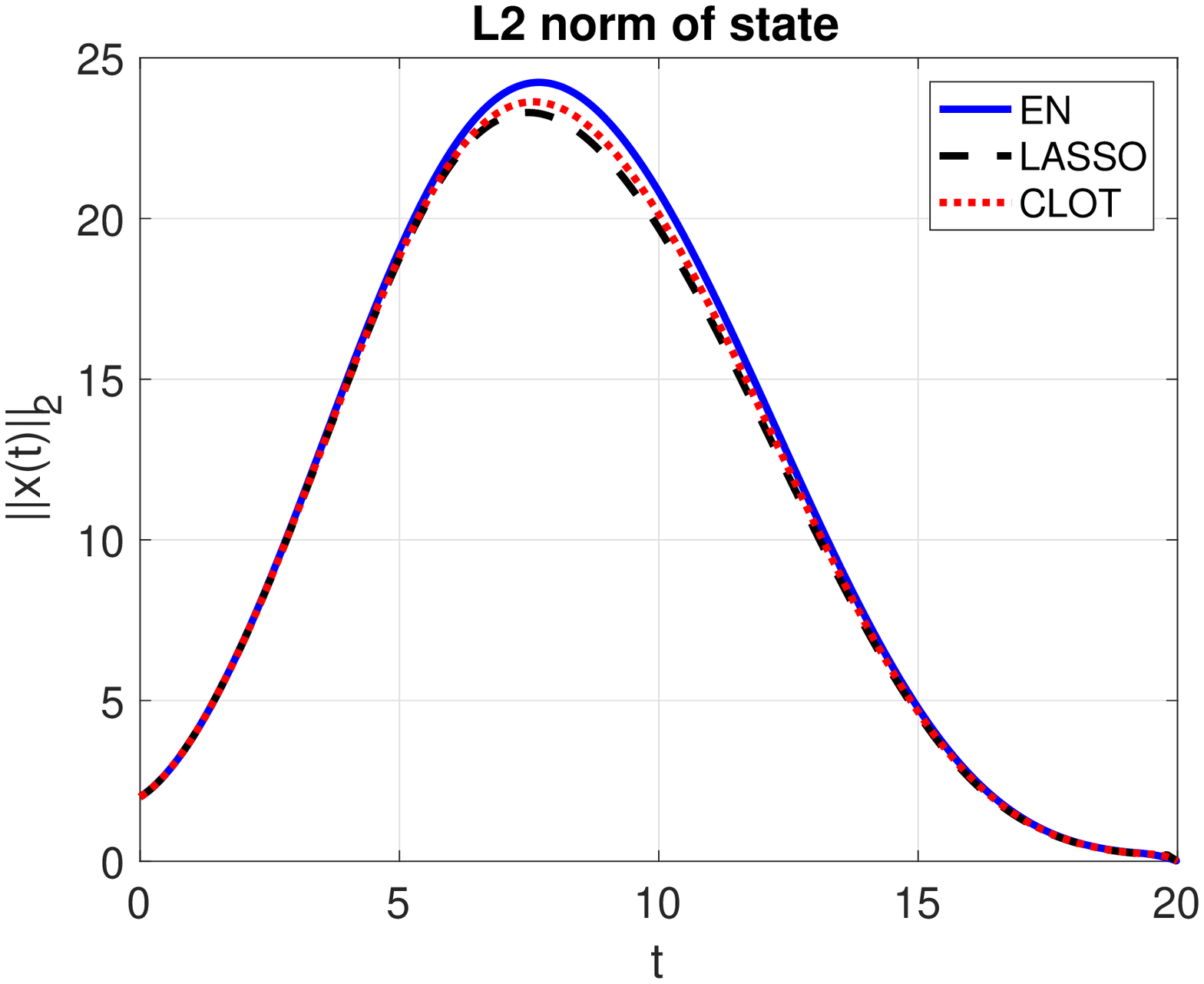}
		\ec
\caption{State trajectory for the plant $P_1(s)$ with the
initial state $(1,1,1,1)^\top$ and $\l = 1$.}
\label{fig:P1_state_l_1}
\efig

\bfig
		\bc
		\includegraphics[width=80mm]{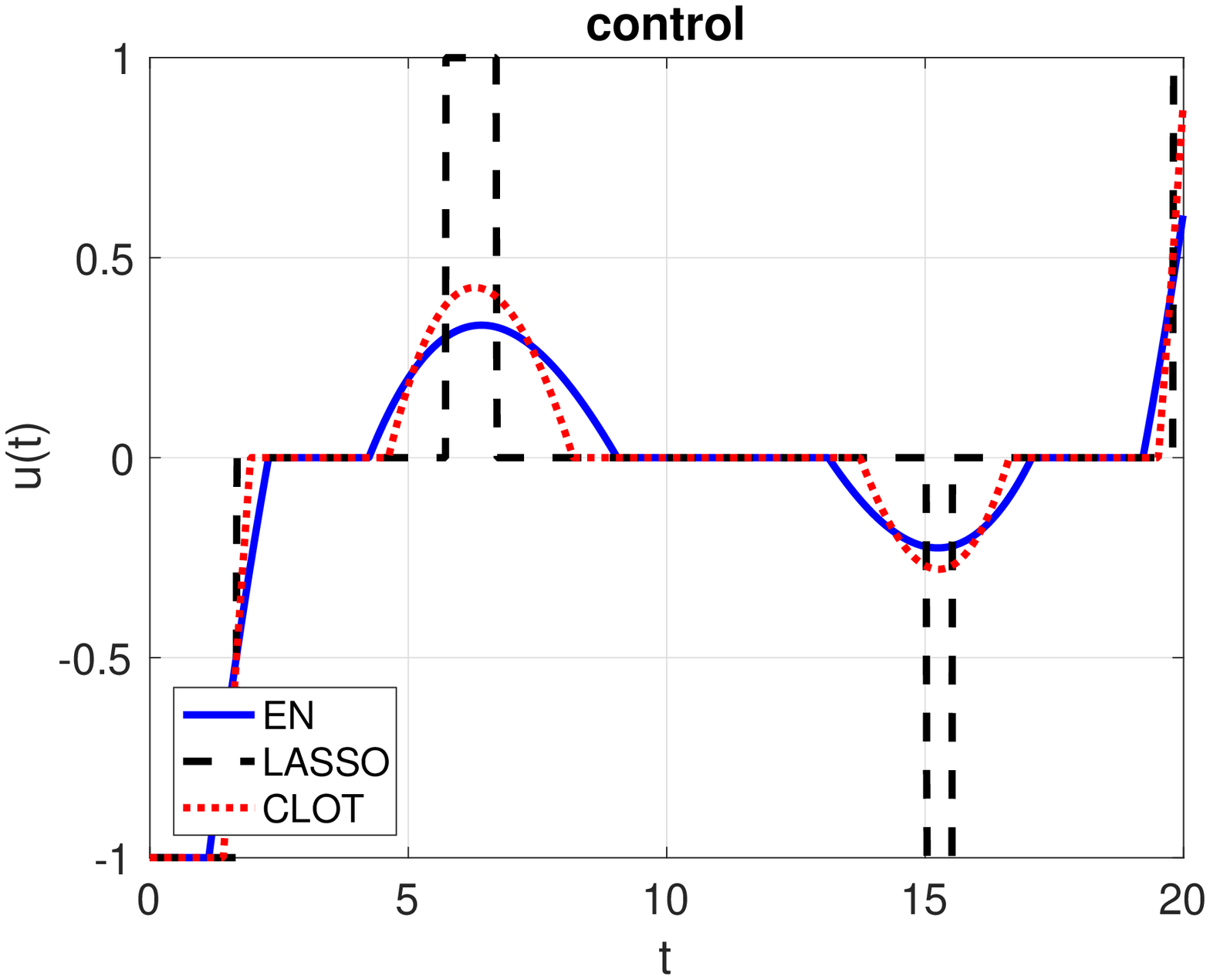}
		\ec
	\caption{Control trajectory for the plant $P_1(s)$ with the
initial state $(1,1,1,1)^\top$ and $\l = 1$.}
\label{fig:P1_control_l_1}
	\efig

		\bfig
		\bc
		\includegraphics[width=80mm]{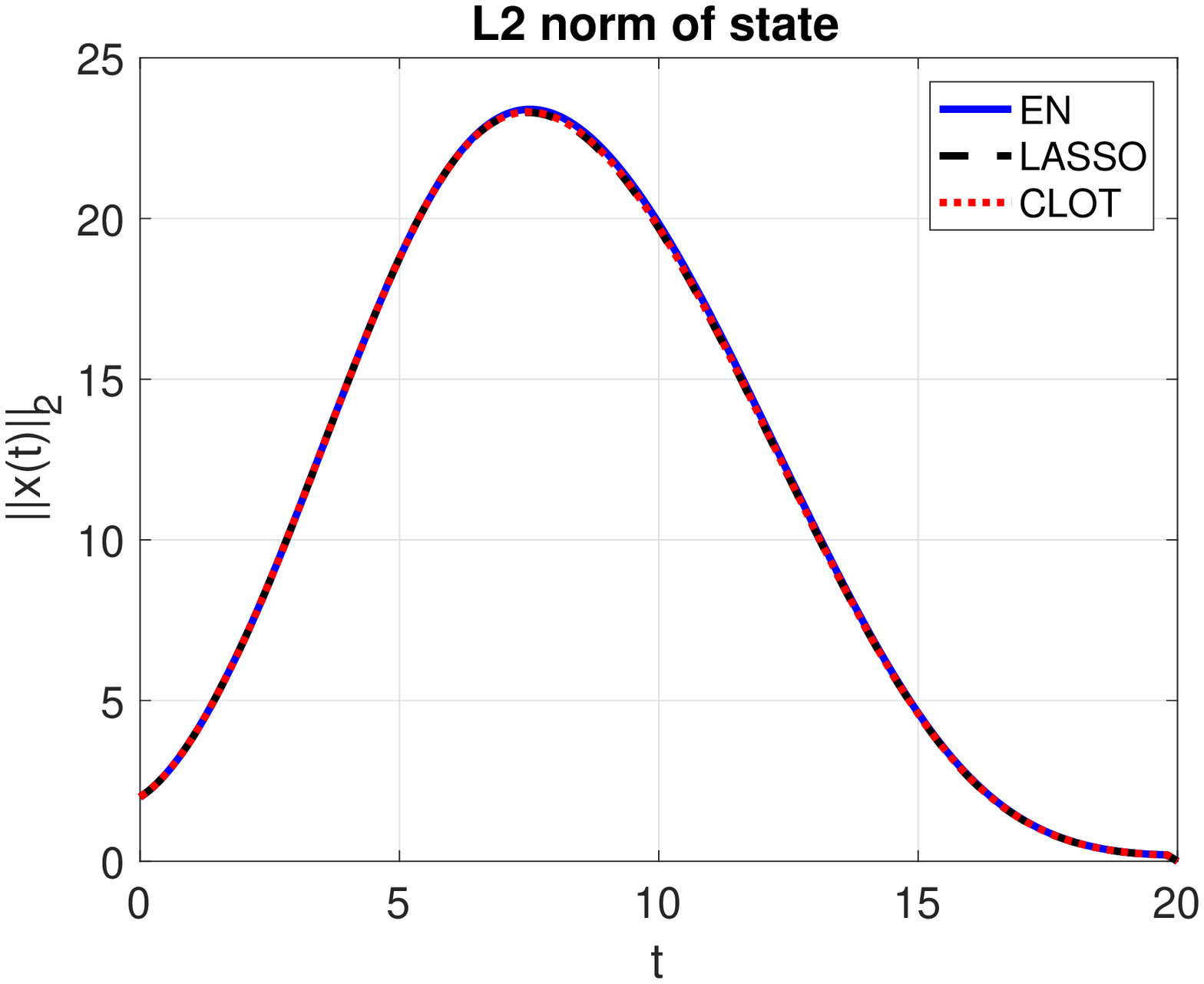}
		\ec
\caption{State trajectory for the plant $P_1(s)$ with the
initial state $(1,1,1,1)^\top$ and $\l = 0.1$.}
\label{fig:P1_state_l_01}
\efig

\bfig
		\bc
		\includegraphics[width=80mm]{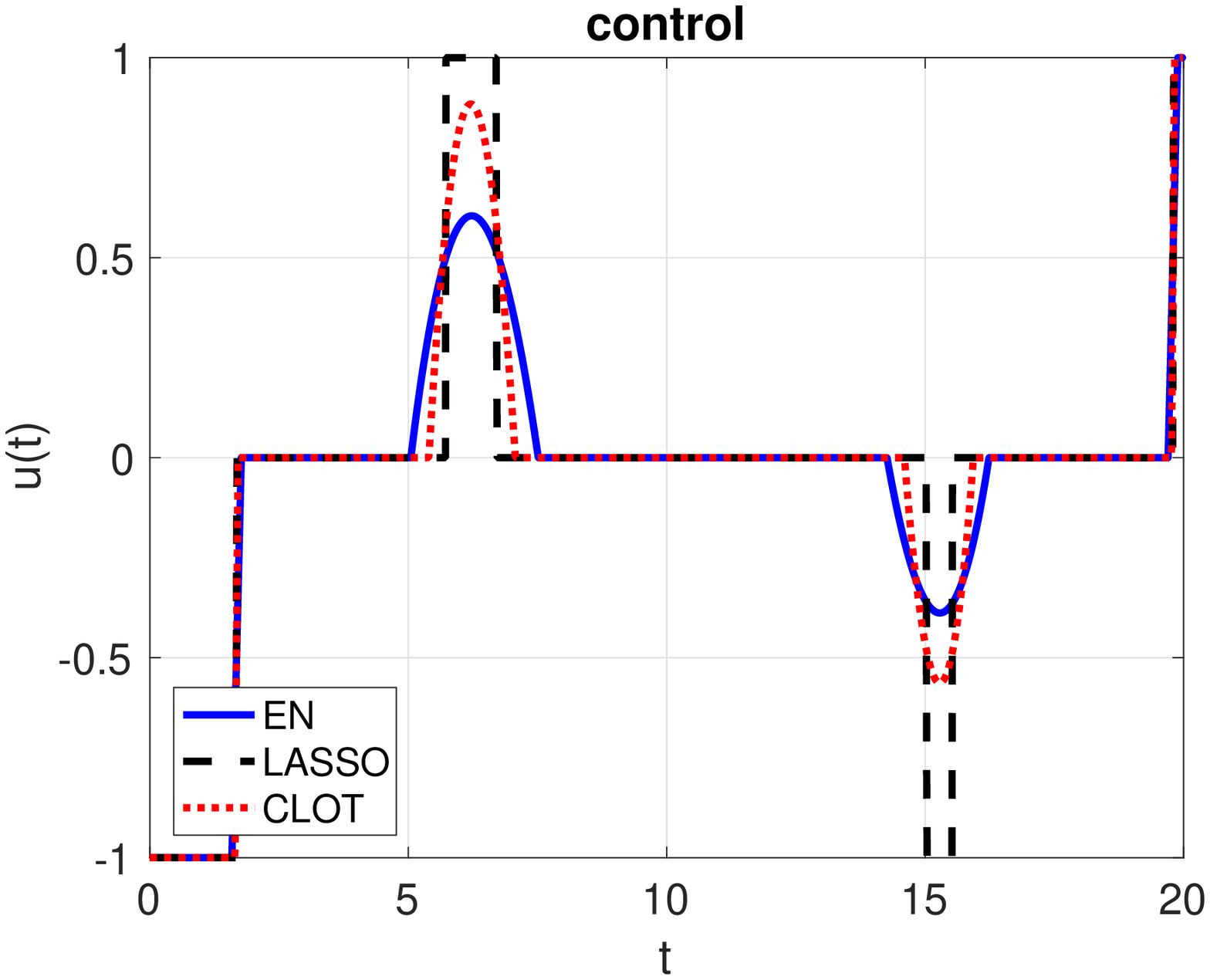}
		\ec
\caption{Control trajectory for the plant $P_1(s)$ with the
initial state $(1,1,1,1)^\top$ and $\l = 0.1$.}
	\label{fig:P1_control_l_01}
	\efig
	
Figures \ref{fig:P2_state_1_1} and \ref{fig:P2_control_1_1}
display the state trajectory and the control
trajectories of the plant $P_2(s)$ (damped harmonic oscillator)
when the initial state is $(1,1)^\top$.
Figures \ref{fig:P2_state_10_1} and \ref{fig:P2_control_10_1}
show the state and control trajectories with the initial state
$(10,1)^\top$.
It can be seen that, with this intial state, the control signal
changes sign more frequently.

	\bfig
		\bc
		\includegraphics[width=80mm]{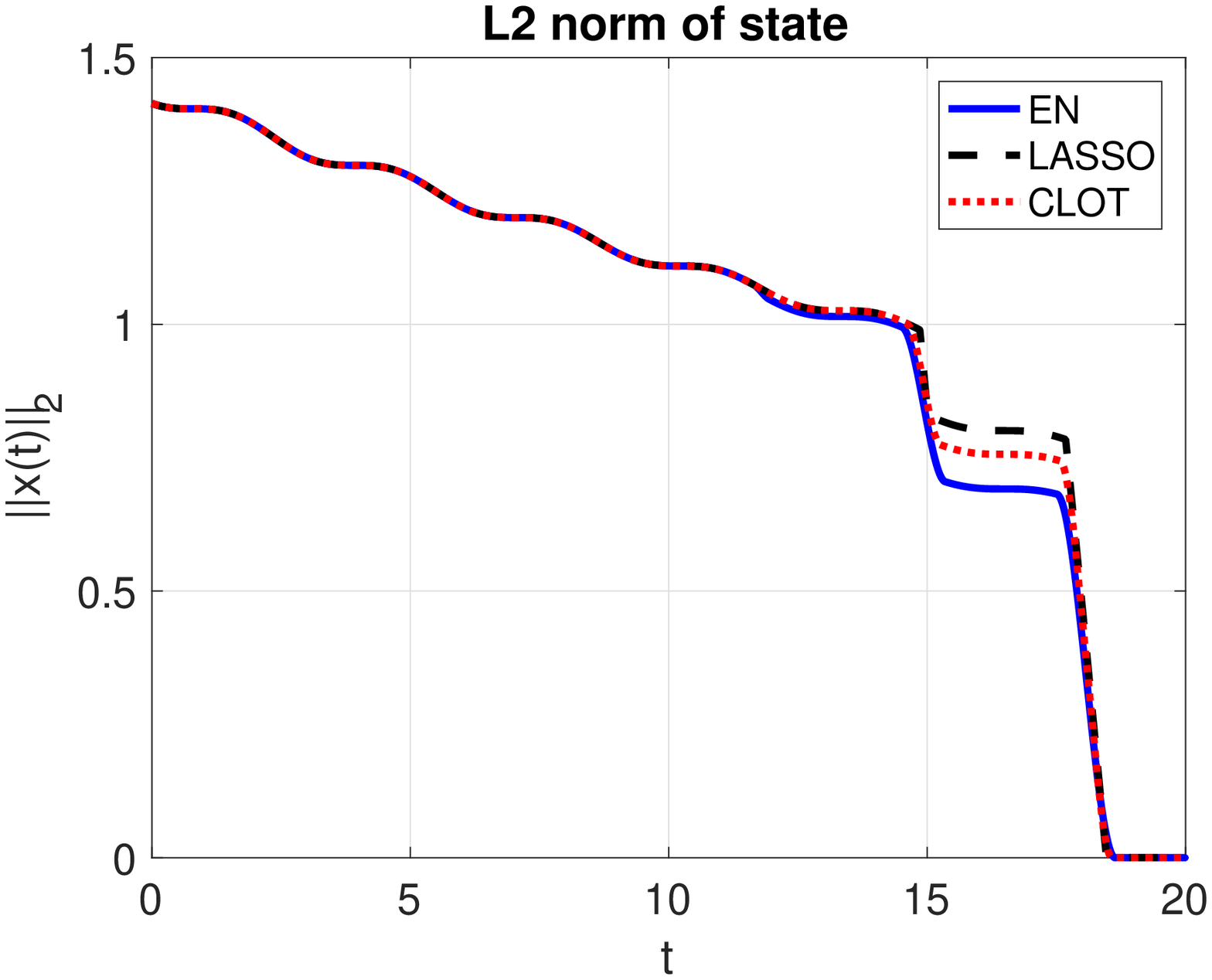}
		\ec
\caption{State trajectory for the the plant $P_2(s)$ with the
initial state $(1,1)^\top$ and $\l = 0.1$.}
\label{fig:P2_state_1_1}
\efig

\bfig
		\bc
		\includegraphics[width=80mm]{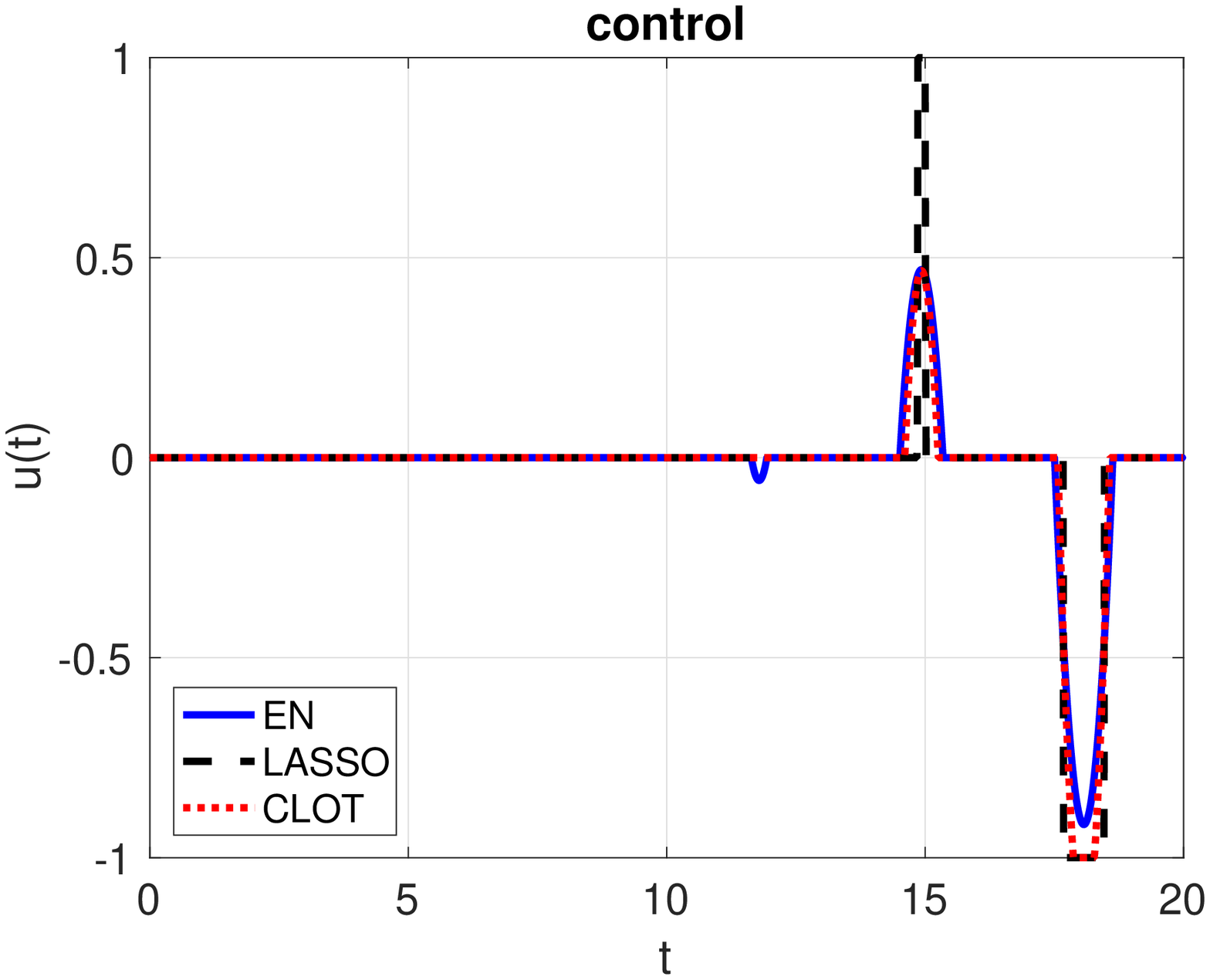}
		\ec
\caption{Control trajectory for the the plant $P_2(s)$ with the
initial state $(1,1)^\top$ and $\l = 0.1$.}
\label{fig:P2_control_1_1}
	\efig

\bfig
\bc
\includegraphics[width=80mm]{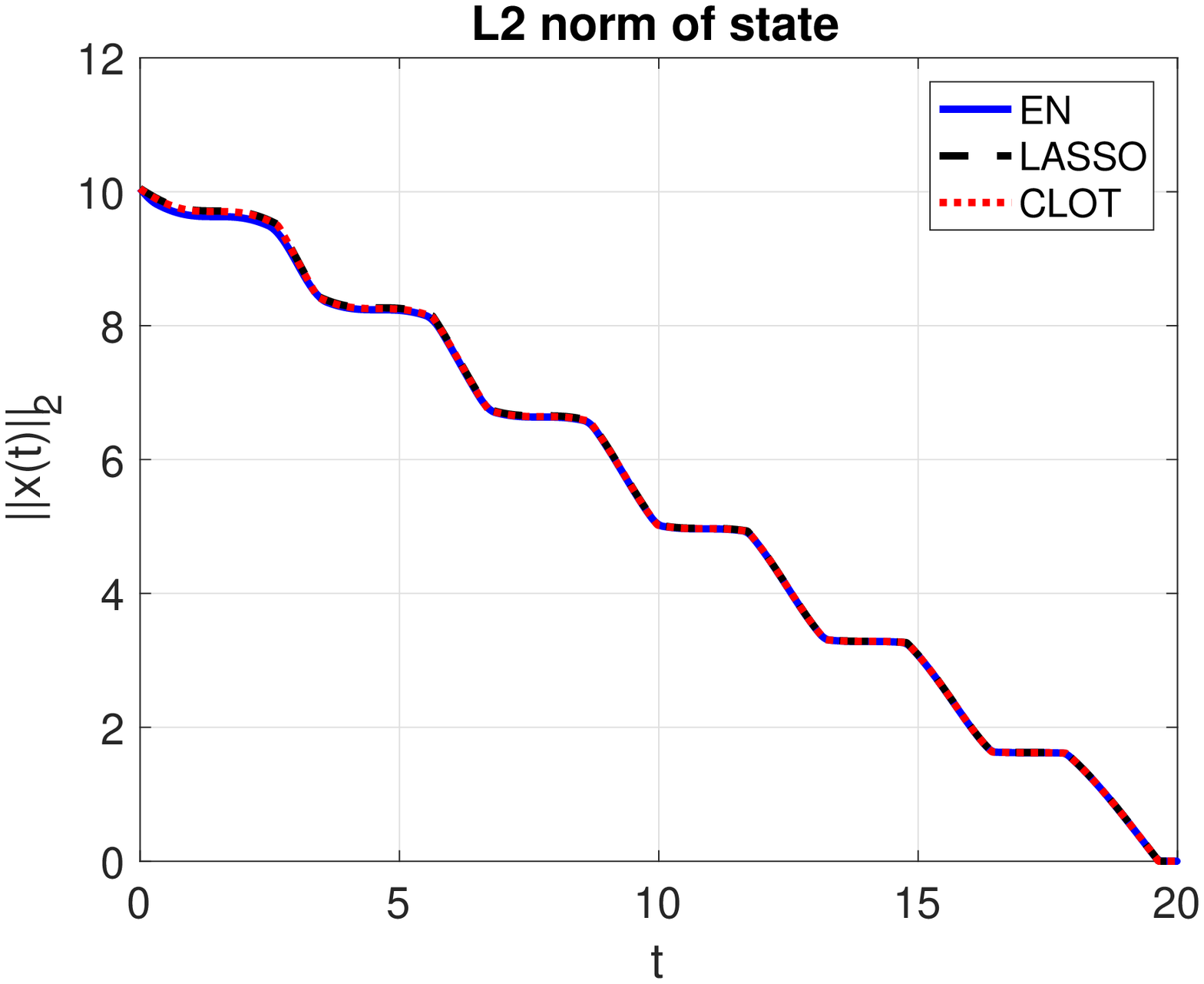}
\ec
\caption{State trajectory for the plant $P_2(s)$ with the
initial state $(10,1)^\top$ and $\l = 0.1$.}
\label{fig:P2_state_10_1}
\efig

\bfig
\bc
\includegraphics[width = 80mm]{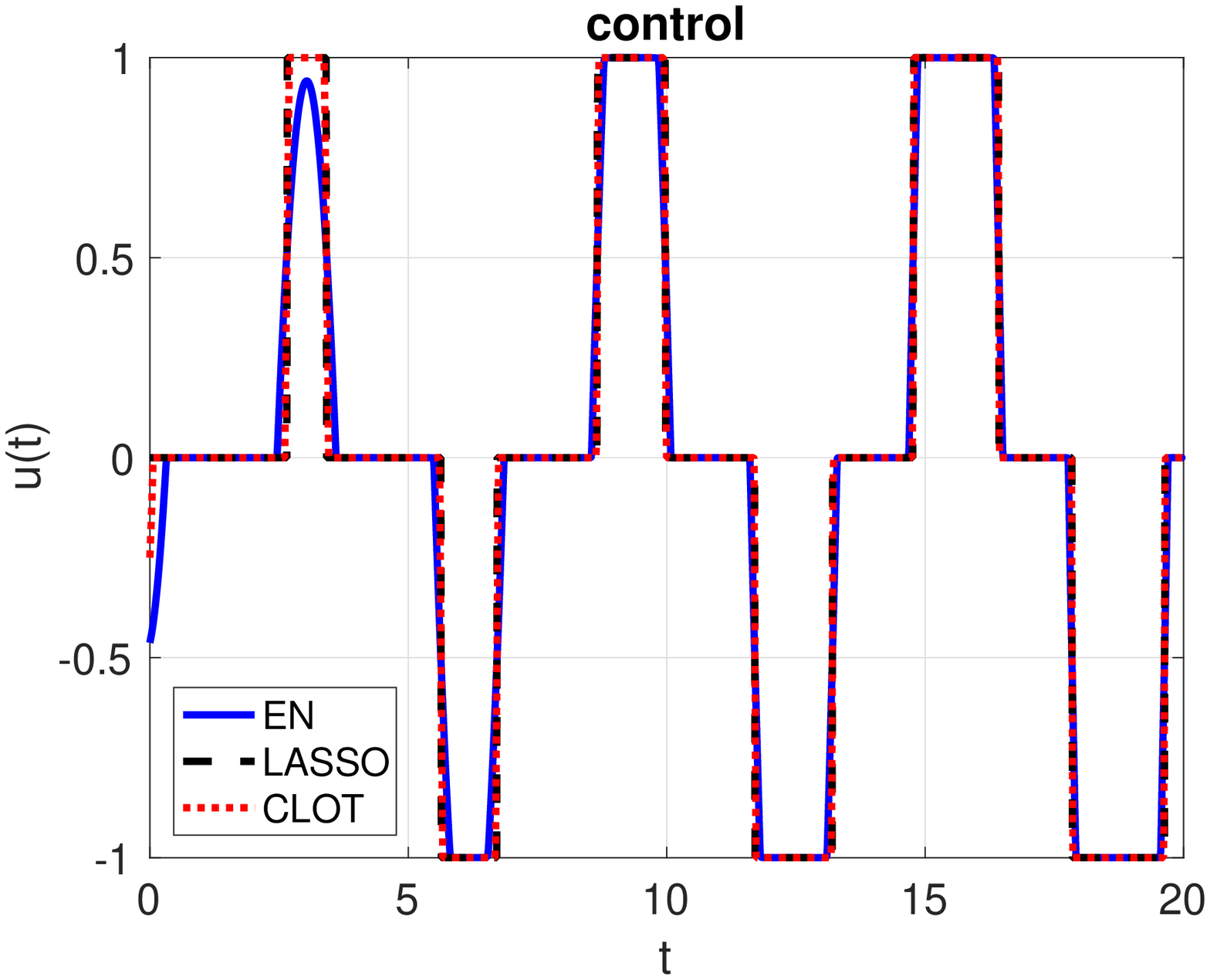}
\ec
\caption{Control trajectory for the plant $P_2(s)$ with the
initial state $(10,1)^\top$ and $\l = 0.1$.}
\label{fig:P2_control_10_1}
\efig

\bfig
\bc
\includegraphics[width=80mm]{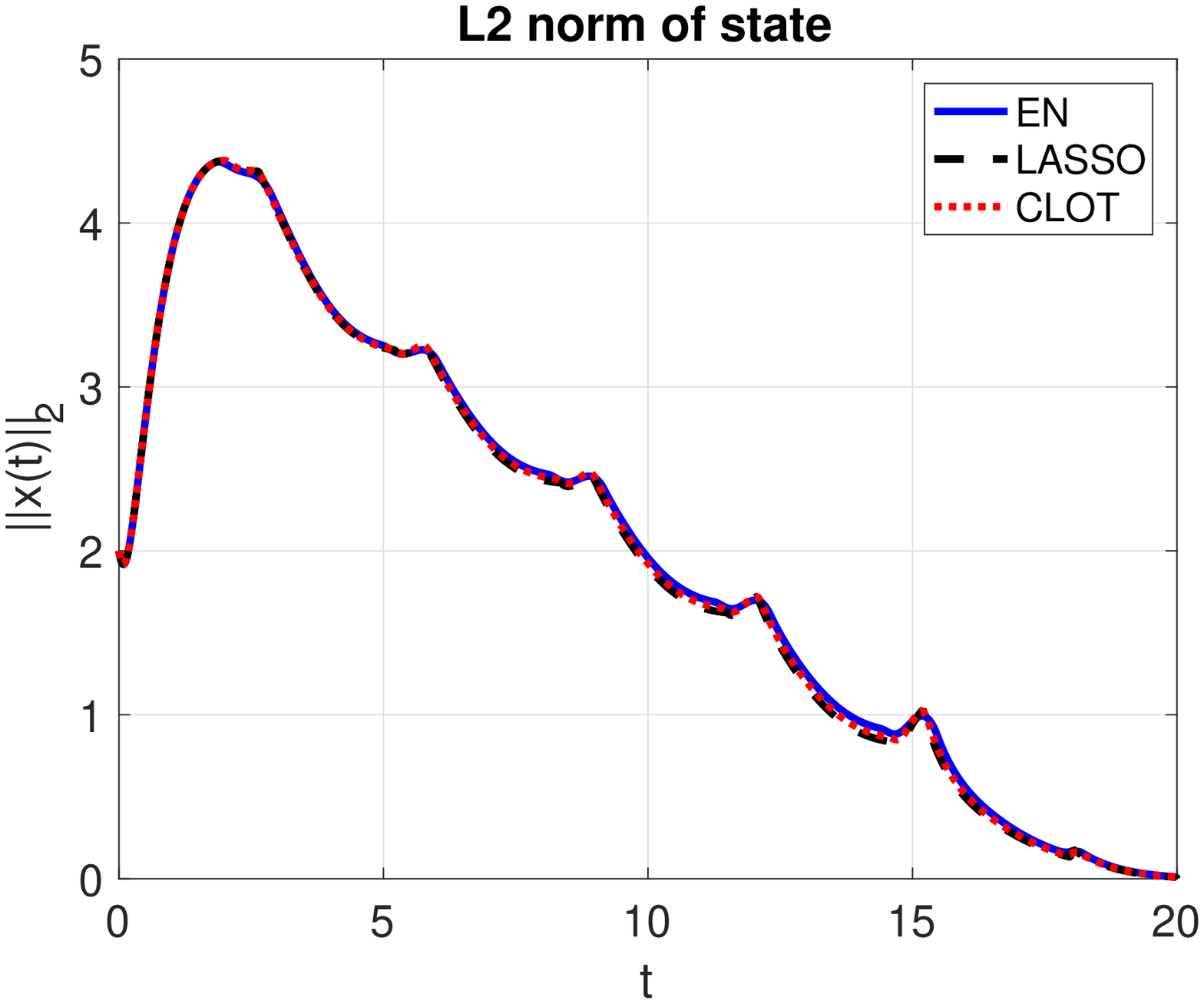}
\ec
\caption{State trajectory for the plant $P_3(s)$ with the initial
state $(1,1,1,1)^\top$ and $\l = 0.1$.}
\label{fig:P3_state}
\efig

\bfig
\bc
\includegraphics[width=80mm]{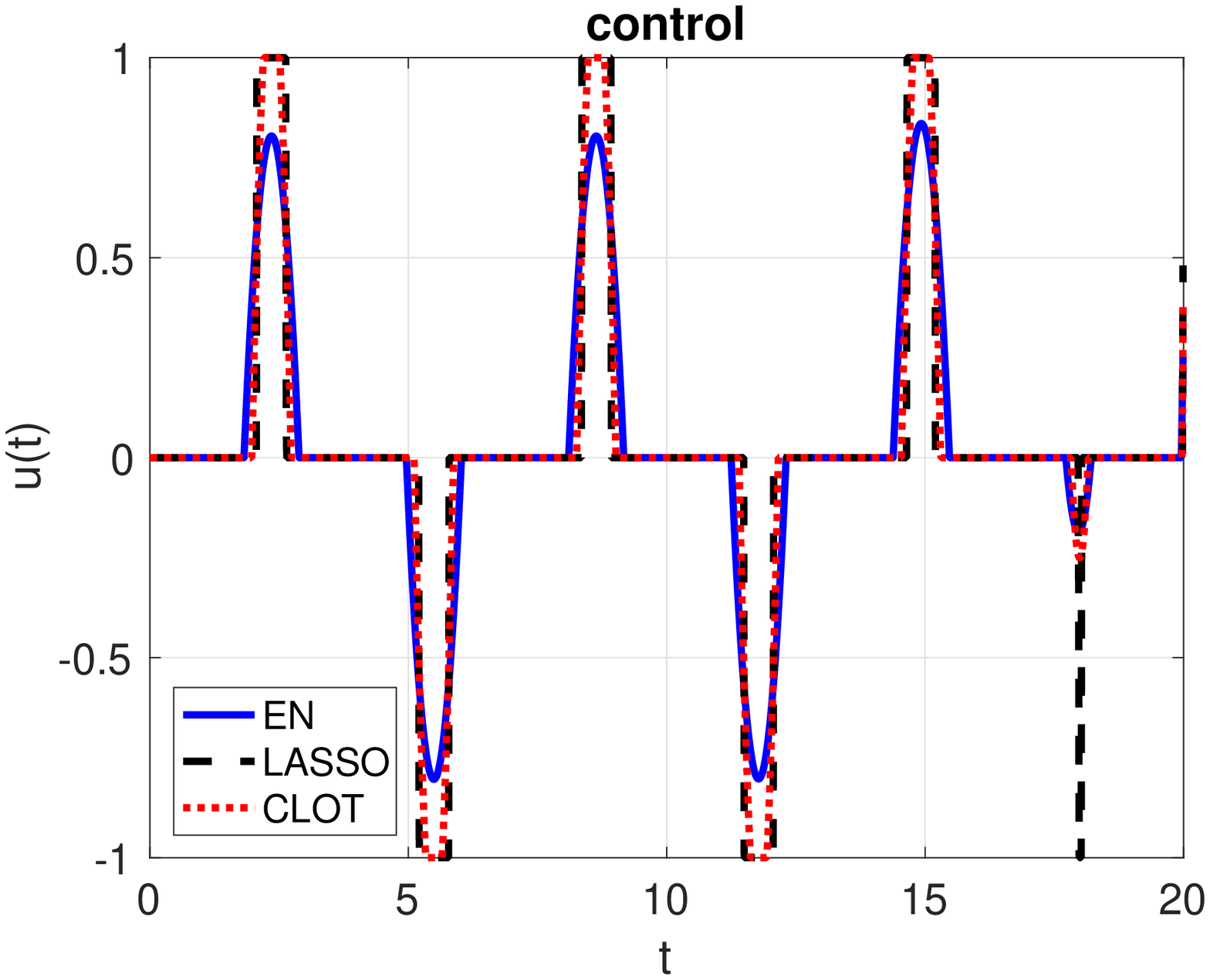}
\ec
\caption{Control trajectory for the plant $P_3(s)$ with the initial
state $(1,1,1,1)^\top$ and $\l = 0.1$.}
\label{fig:P3_control}
\efig

%

\bfig
\bc
\includegraphics[width=80mm]{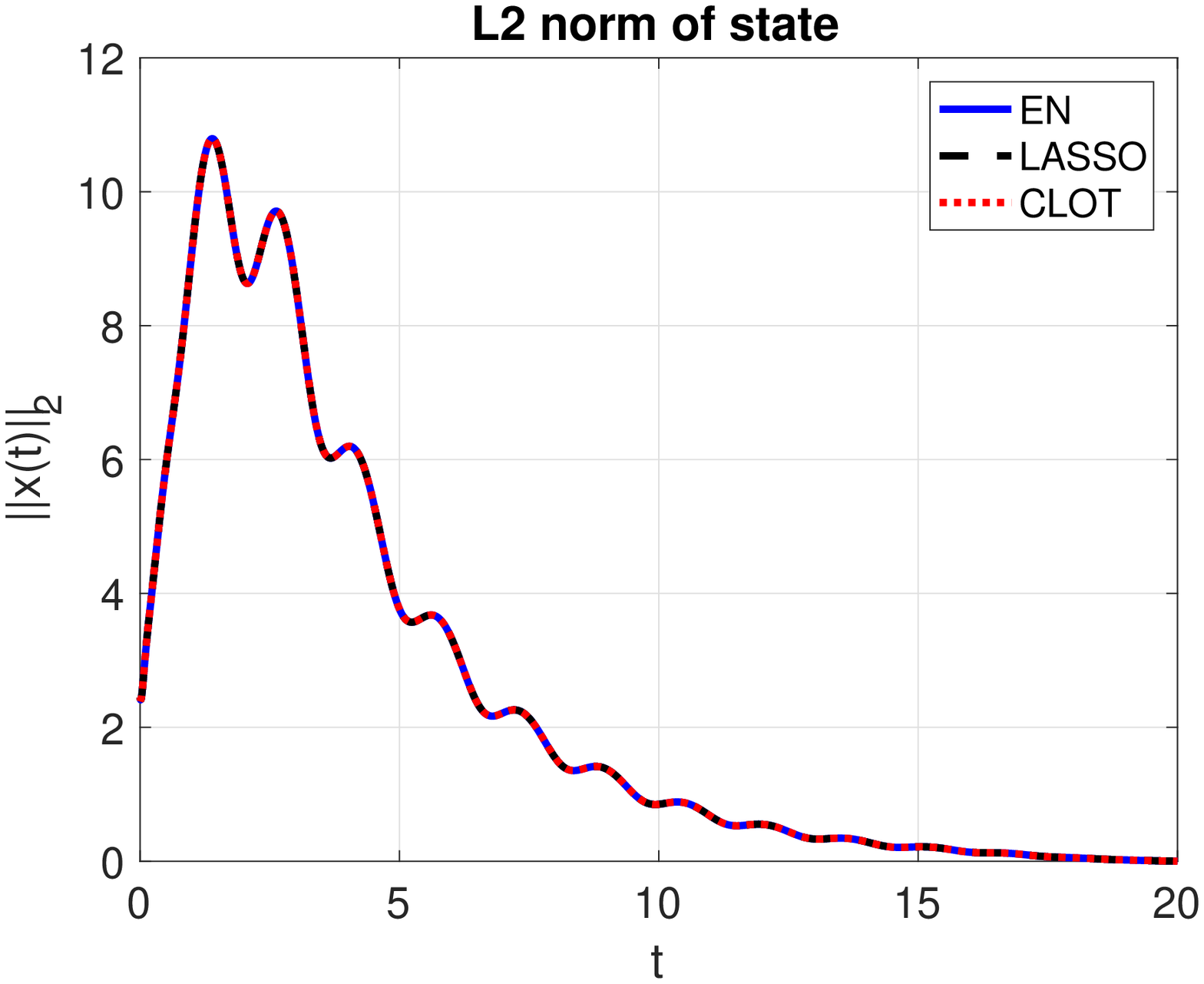}
\ec
\caption{State trajectory for the plant $P_4(s)$ with the initial
state $(1,1,1,1,1,1)^\top$ and $\l = 0.1$.}
\label{fig:P5_state}
\efig

\bfig
\bc
\includegraphics[width=80mm]{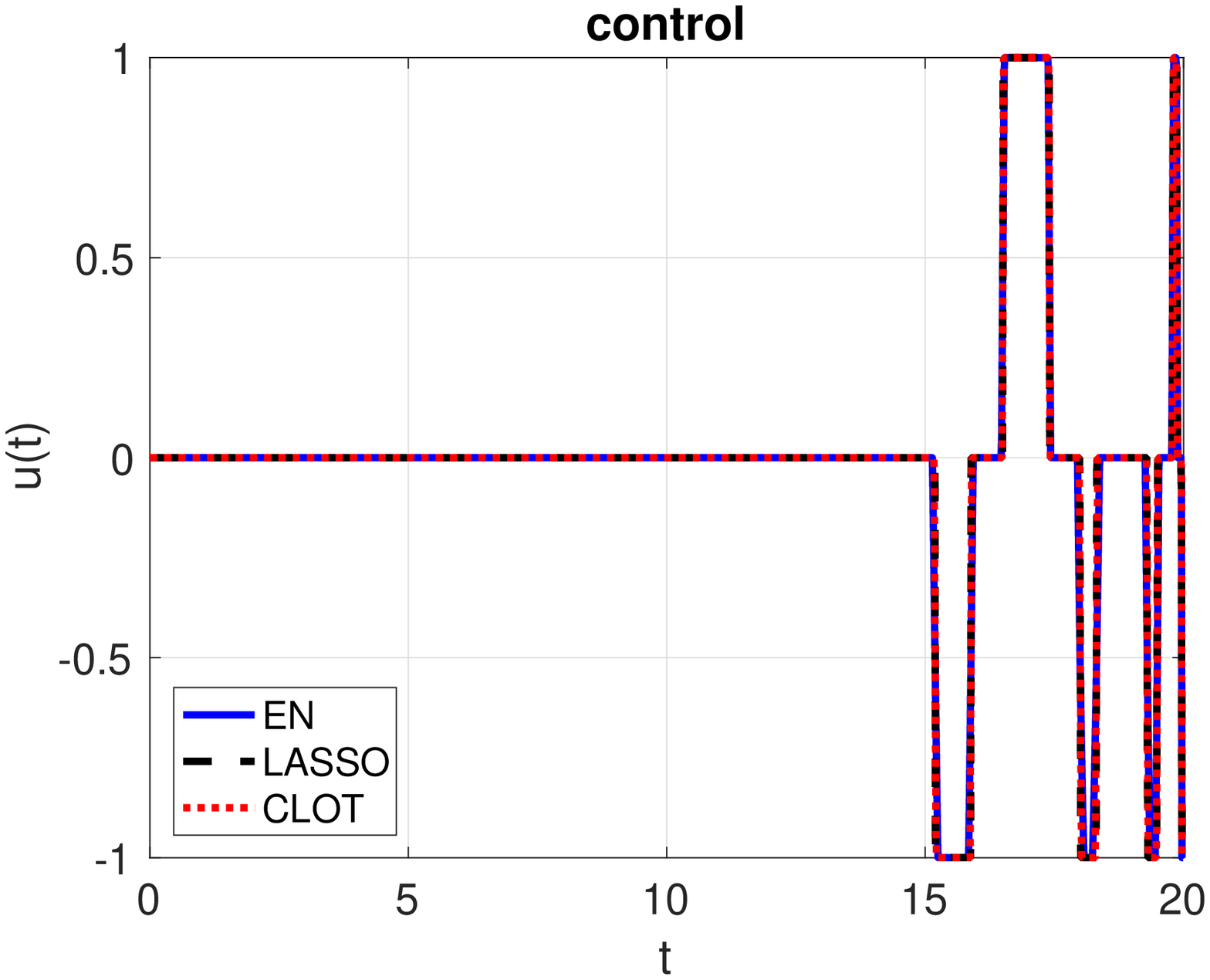}
\ec
\caption{Control trajectory for the plant $P_4(s)$ with the initial
state $(1,1,1,1,1,1)^\top$ and $\l = 0.1$.}
\label{fig:P5_control}
\efig

\bfig
\bc
\includegraphics[width=80mm]{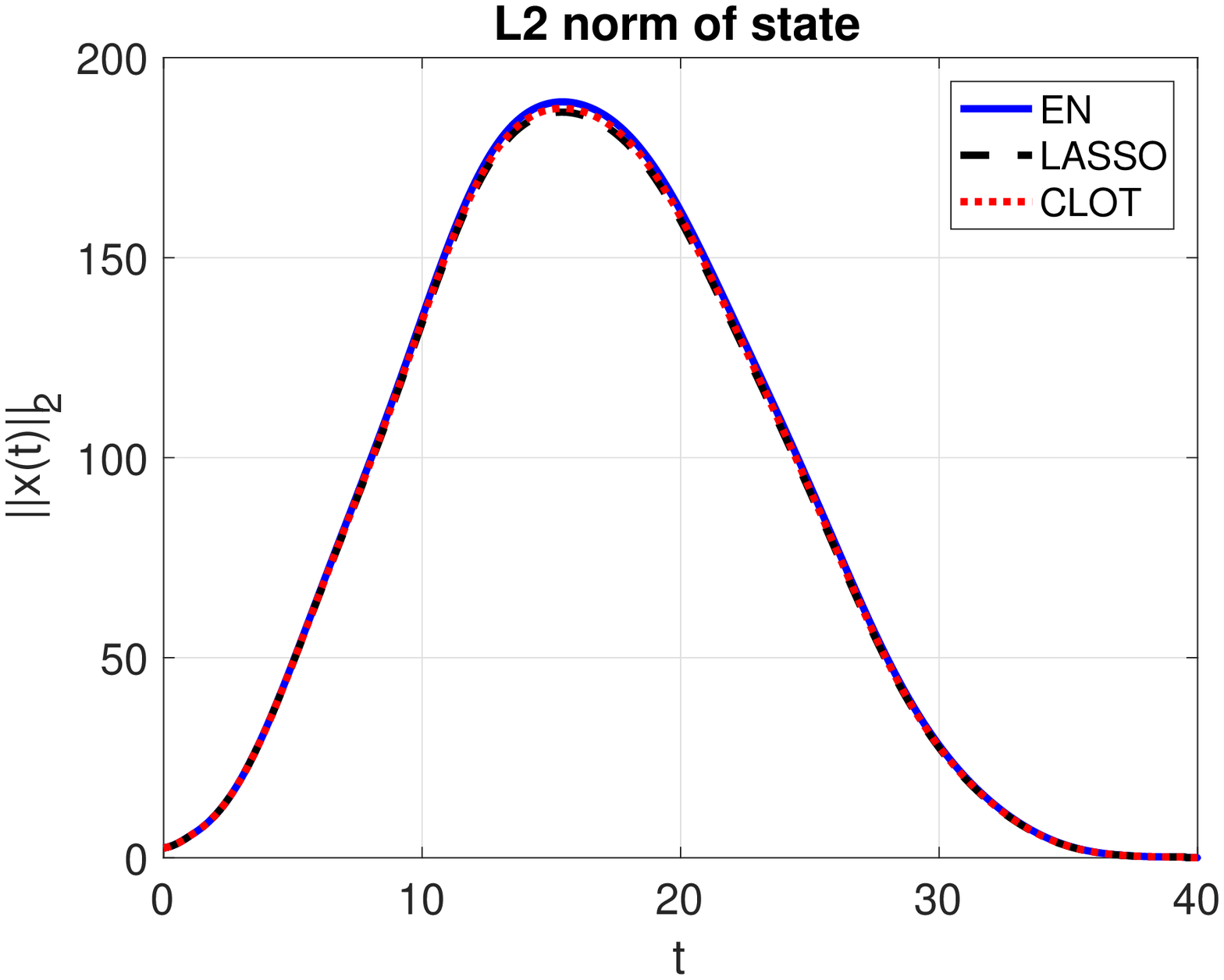}
\ec
\caption{State trajectory for the plant $P_5(s)$ with the initial
state $(1,1,1,1,1,1)^\top$ and $\l = 0.1$.}
\label{fig:P6_state}
\efig

\bfig
\bc
\includegraphics[width=80mm]{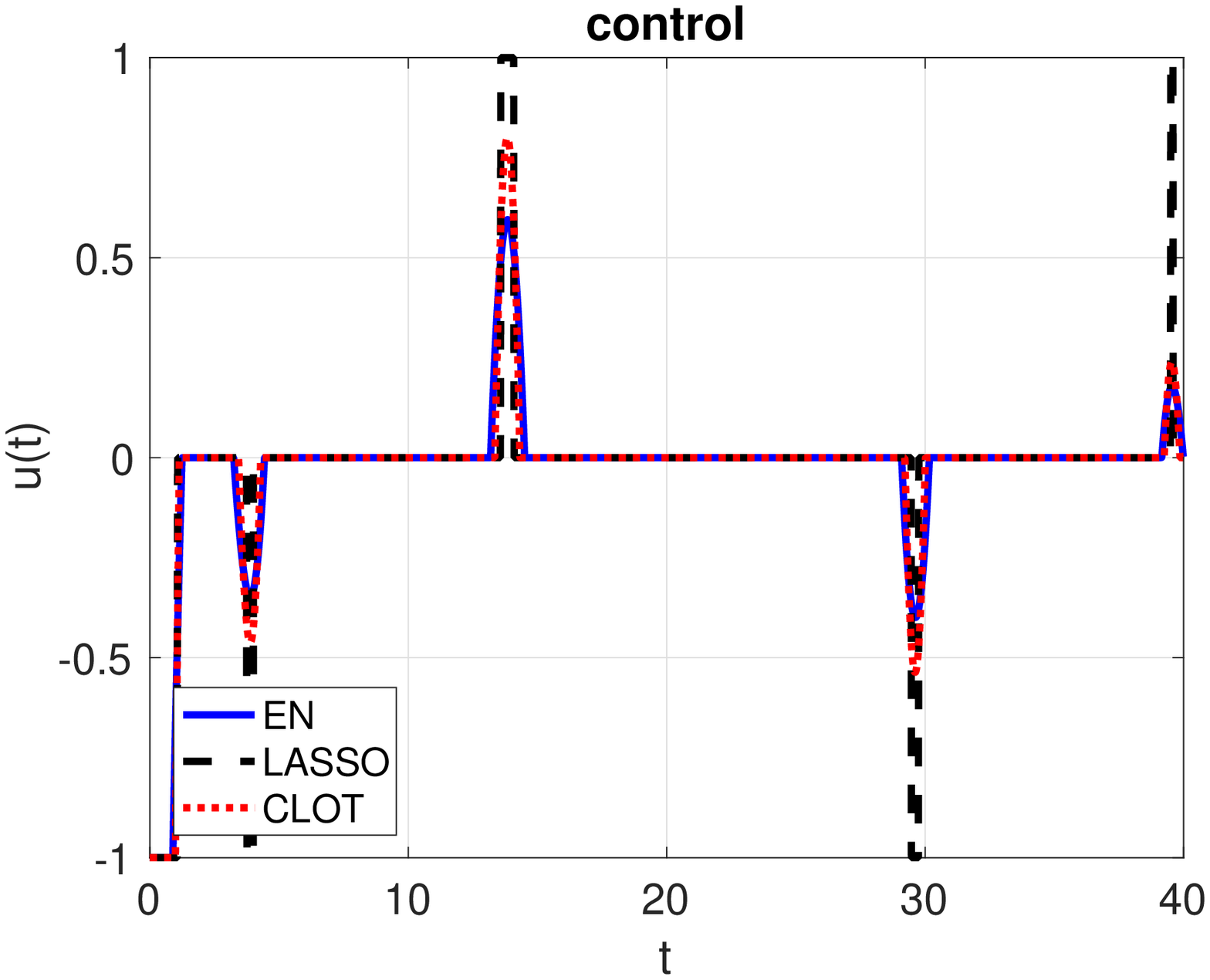}
\ec
\caption{Control trajectory for the plant $P_5(s)$ with the initial
state $(1,1,1,1,1,1)^\top$ and $\l = 0.1$.}
\label{fig:P6_control}
\efig

\bfig
\bc
\includegraphics[width=80mm]{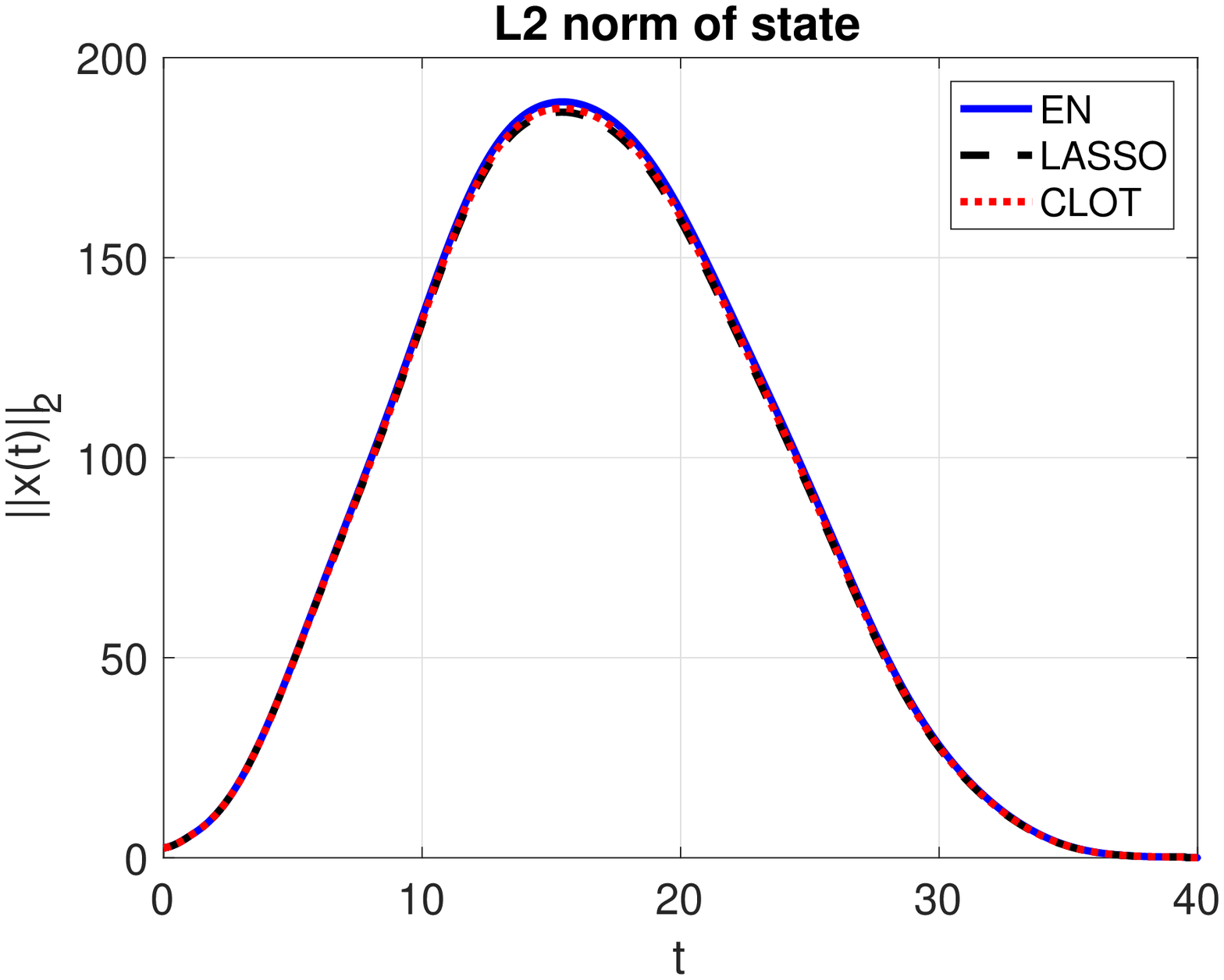}
\ec
\caption{State trajectory for the plant $P_6(s)$ with the initial
state $(1,1,1,1,1,1)^\top$ and $\l = 0.1$.}
\label{fig:P7_state}
\efig

\bfig
\bc
\includegraphics[width=80mm]{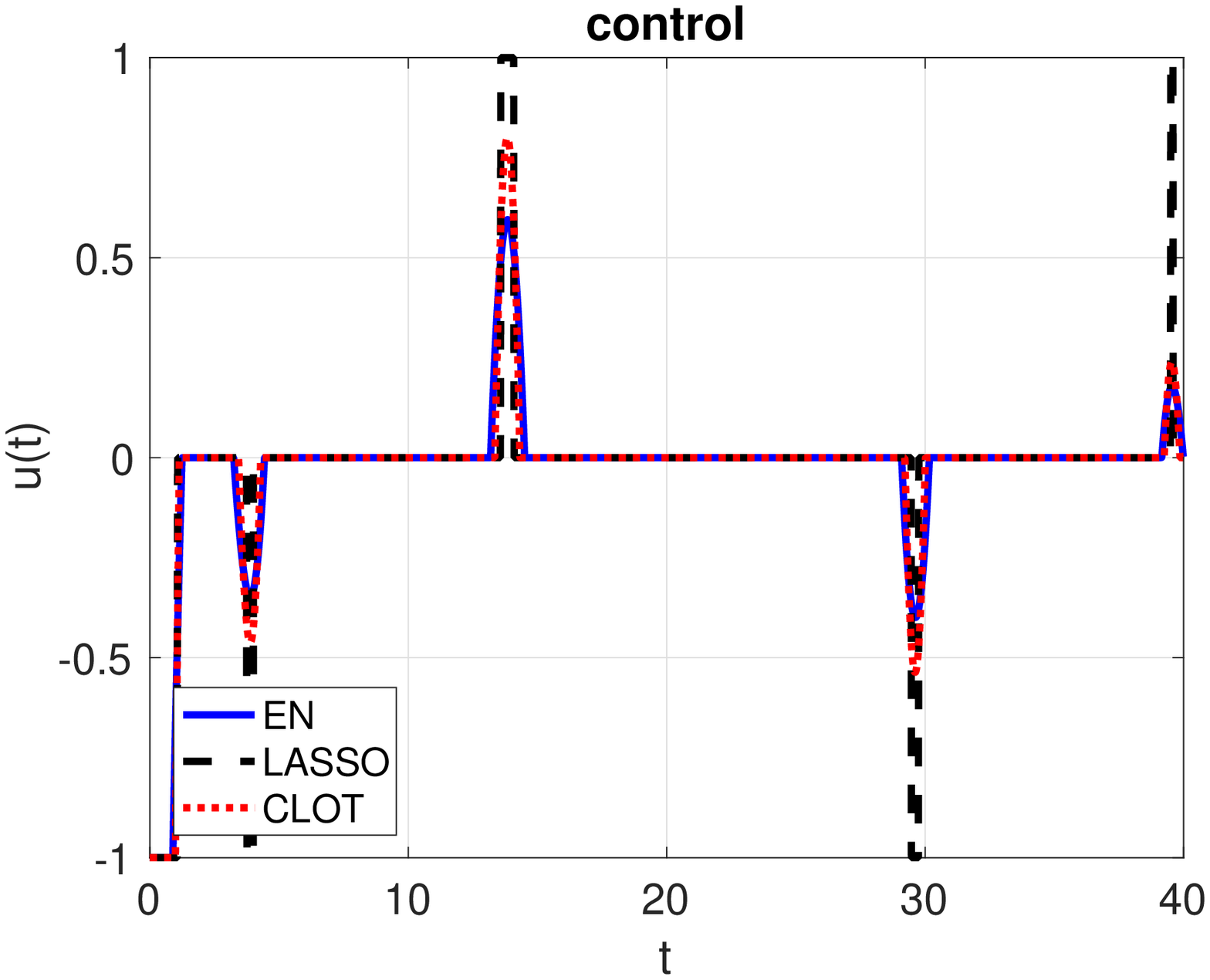}
\ec
\caption{Control trajectory for the plant $P_6(s)$ with the initial
state $(1,1,1,1,1,1)^\top$ and $\l = 0.1$.}
\label{fig:P7_control}
\efig

To compare the sparsity densities of the three control signals,
we compute the fraction of time that each signal is nonzero.
In this connection, it should be noted that the LASSO control signal
is the solution of a \textit{linear programming} problem; consequently
its components \textit{exactly} equal zero at many time instants.
In contrast, the EN and CLOT control signals are the solutions of
\textit{convex} optimization problems.
Consequently, there are many time instants when the control signal
is ``small'' without being smaller than the machine zero.
Therefore, to compute the sparsity density, we applied a threshold of
$10^{-4}$, and treated a component of a control signal as being zero if
its magnitude is smaller than this threshold.
With this convention, the sparsity densities of the various control signals
are as shown in Table \ref{table:int}.
From this table it can be seen that the control signal generated using
CLOT norm minimization has significantly lower sparsity density
compared to that of EN, and is not much higher than that of LASSO.
Also, as expected, the sparsity density of LASSO does not change with $\l$,
whereas the sparsity densities of both EN and CLOT decrease as $\l$ is 
decreased.
For this reason, in other examples we present only the results for $\l = 0.1$.
\begin{table}
\bc
\btab{|l|r|r|r|}
\hline
\textbf{$\l$} & \textbf{LASSO} & \textbf{EN} & \textbf{CLOT} \\
\hline
$\l = 1$ & 0.1725 & 0.6050 & 0.5900 \\
        \hline
$\lambda = 0.1$ & 0.1725 & 0.3795 & 0.2665 \\ 
\hline
\etab
\ec
\caption{Sparsity indices of the control signals from various algorithms
for the plant $P_1(s)$ (fourth-order integrator) with the initial state
$(1,1,1,1)$.}
\label{table:int}
\end{table}
\subsection{Comparison of Sparsity Densities}

In this subsection we analyze the sparsity densities, that is, the
fraction of samples that are nonzero, using the three methods LASSO, EN,
and CLOT.
The advantage of using the sparsity \textit{density} instead of the sparsity
\textit{count} (the absolute number of nonzero entries) is that when
the sample time is reduced, the sparsity count would increase, whereas
we would expect the sparsity density to remain the same.
As explained above, we have applied a threshold of $10^{-4}$ in computing the
sparsity densities of various control signals.

Table \ref{table:spars} shows the sparsity densities for the nine
examples studied in Table \ref{table:plants}, in the same order.
From this table it can be seen that the CLOT norm-based control signal
is always more sparse than the EN-based control signal.
Indeed, in some cases the sparsity density of the CLOT control is
comparable to that of the LASSO control.                                                   
\begin{table}		
\bc
\btab{|c|c|c|c|}
\hline
\textbf{No.} & \textbf{LASSO} & \textbf{EN} & \textbf{CLOT}\\
\hline
1 & 0.1690 & 0.5915 & 0.4450 \\
\hline
2 & 0.1690 & 0.3250 & 0.2535 \\ \hline
3 & 0.0480 & 0.1130 & 0.0830 \\ \hline
4 & 0.4055 & 0.5560 & 0.4225 \\ \hline
5 & 0.1460 & 0.2935 & 0.2075 \\ \hline
6 & 0.1125 & 0.1310 & 0.1175 \\ \hline
7 & 0.0568 & 0.1490 & 0.1125 \\ \hline
8 & 0.0568 & 0.1490 & 0.1125 \\ \hline
\etab
\ec
\caption{Sparsity densities for optimal controllers produced by
various methods}
\label{table:spars}
\end{table}

We also increased the number of samples from 2,000 to 4,000, and the
optimal values changed only in the third significant figure in almost
all examples for all three methods.
Therefore the figures in Table \ref{table:spars} are essentially equal to
the Lebesgue measure of the support set divided by $T$.
\section{Numerical Examples with state constraints}\label{sec:exam_state}

In this section we present numerical results from applying the CLOT norm
minimization approach to two different plants imposed with state constraints on a range of thresholds ($\theta$), and compare the results with those from applying LASSO and EN.

\subsection{Details of the plants}

The plants $P_1(s)$ and $P_7(s)$ defined in the table \ref{table:plants} are used to demonstrated the results with state constraints. The parameters for each plant used for the optimization problems are listed in the table \ref{table:plants_state}.

\begin{table}
	\bc
	\btab{|c|c|c|c|c|}
	\hline
	\textbf{Plant} & $T$ & $x(0)$ & $\l$ & Range of $\theta$ \\
	\hline
	$P_1(s)$ & 20 & $[1,0,1,1]^\top$ & 1  & (6, 10)\\
	\hline
	$P_7(s)$ & 40 & $[1,0,1,1,1,1]^\top$ & 0.1 & (30, 200) \\
	\hline
	\etab
	\ec
	\label{table:plants_state}
	\caption{Details of various plants studied under state constraints}
\end{table}
\subsection{Comparison of Sparsity Densities}

In this subsection we analyze the sparsity densities, using the three methods LASSO, EN, and CLOT across the range of $\theta$ mentioned in the table \ref{table:plants_state}.

Figure \ref{fig:P1_comp_sc} shows the sparsity densities of the plant $P_1(s)$ w.r.t. the threshold $\theta$, where the threshold is increased in steps of 0.5.Figure \ref{fig:P7_comp_sc} shows the sparsity densities of the plant $P_7(s)$ w.r.t. the threshold $\theta$, where the threshold is increased in steps of 1. Figure \ref{fig:P1_11_state} shows the trajectory of $\ell_2$ norm of the state and figure \ref{fig:P1_11_control} shows the control trajectory of the plant $P_1(s)$ at intial state $[1,0,1,1]^\top$ with $\theta = 11
$.

\bfig
\bc
\includegraphics[width=80mm]{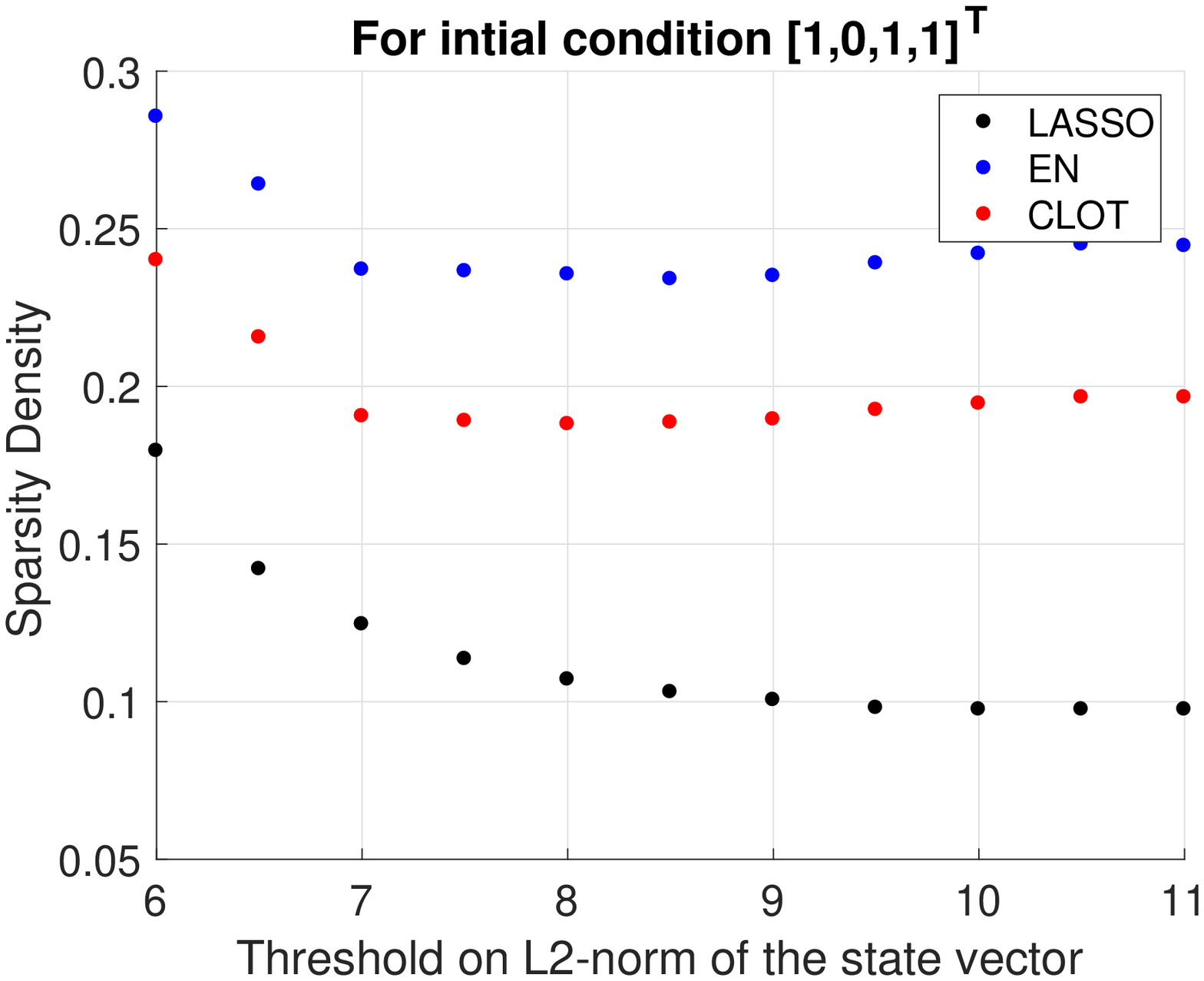}
\ec
\caption{ State threshold  vs Sparsity Density for plant $P_1(s)$ with intial condition $[1,0,1,1]^\top$ }
\label{fig:P1_comp_sc}
\efig

\bfig
\bc
\includegraphics[width=80mm]{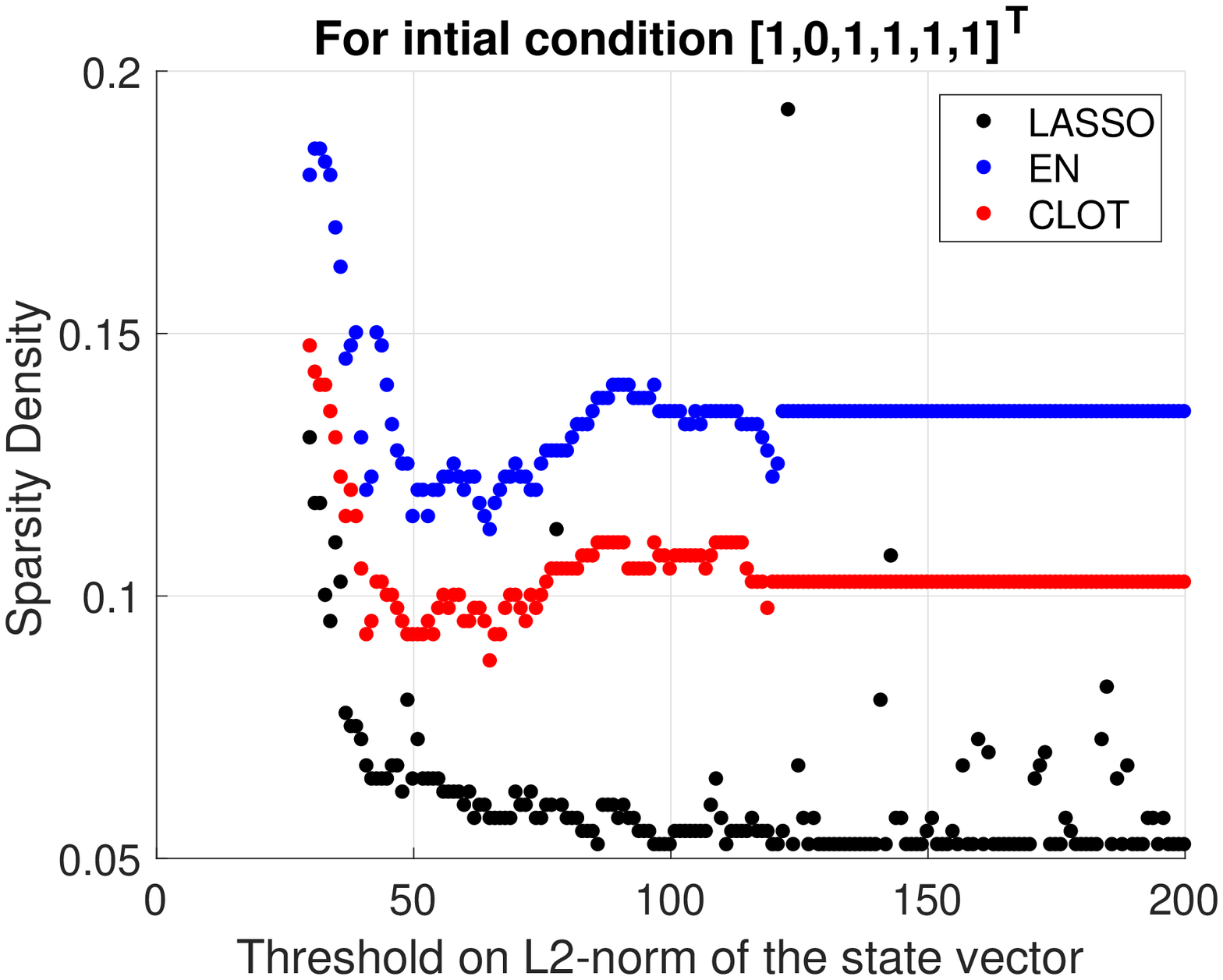}
\ec
\caption{ State threshold  vs Sparsity Density for plant $P_7(s)$ with intial condition $[1,0,1,1,1,1]^\top$ }
\label{fig:P7_comp_sc}
\efig

\bfig
\bc
\includegraphics[width=80mm]{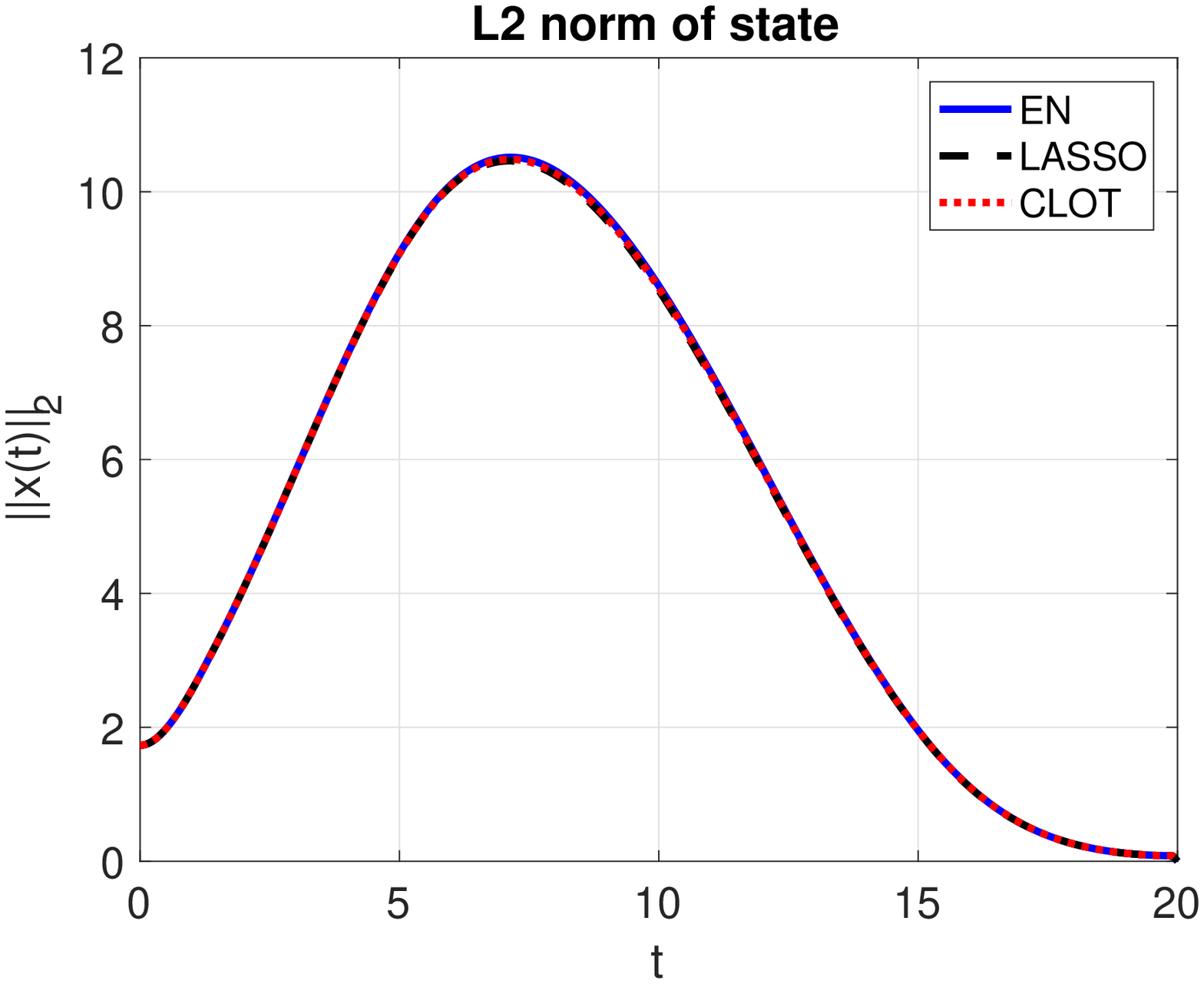}
\ec
\caption{State trajectory for the plant $P_1(s)$ with the initial
	state $(1,0,1,1)^\top$ and $\theta = 11$.}
\label{fig:P1_11_state}
\efig

\bfig
\bc
\includegraphics[width=80mm]{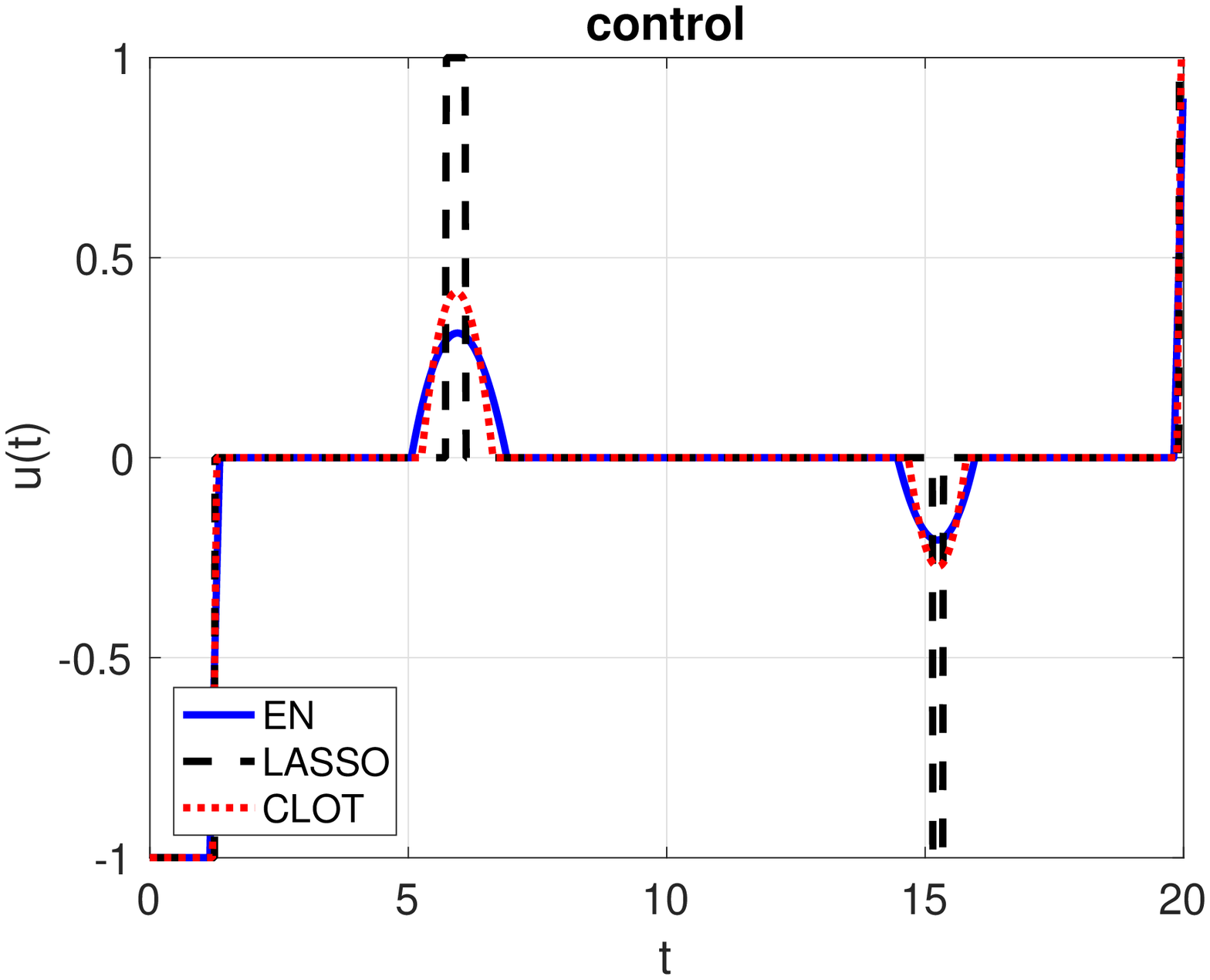}
\ec
\caption{Control trajectory for the plant $P_1(s)$ with the initial
	state $(1,0,1,1)^\top$ and $\theta = 11$.}
\label{fig:P1_11_control}
\efig

There are few points in figure \ref{fig:P7_comp_sc} such as $\theta$ = 123, 143 etc. where Lasso fails to converge when done by cvx package but not EN and CLOT. So, at these points Lasso has higher values for sparsity density. And, for values of $\theta$ around 130 ($\theta_{max}$)  onwards, the sparsity density does not change because at these points, the control input is same.

From figures \ref{fig:P1_comp_sc} and \ref{fig:P7_comp_sc}, it is clearly noted that CLOT control input is more sparse than than that of EN and less sparse compared to that of LASSO.

\section{Conclusions}
\label{sec:Conclusions}

In this article, we propose the CLOT norm-based control that minimizes the
weighted sum of $L^1$ and $L^2$ norms among feasible controls,
to obtain a continuous control signal that is sparser than the EN control
introduced in \cite{NagQueNes16}.
We have shown a discretization method, by which the CLOT optimal control
problem can be solved via finite-dimensional convex optimization.
We have shown that the CLOT control solution is continuous and it approximates $L^1$ optimal control solution. We have also introduced the state constraints to obtain the optimal control, to ensure the states does not blow up in order to get the optimal control.
Numerical experiments have shown the advantage of the CLOT control compared with the LASSO and EN controls.
%
%

\appendix
\section{Preliminaries}\label{app: prelims}

\subsection{Subdifferential of norms \cite{SubgradientMethods:2003wi} } \label{app: sub_diff}
Let $f: \R^n \rightarrow \R \cup \{\infty\}$ where $\mathrm{dom}(f) = \{x \in \R^n: f(x) < \infty\}$. 
A vector $g \in \R^n$ is a \textit{subgradient} of $f$ at some 
$x \in \mathrm{dom}(f)$ if
\begin{equation}\label{eq: subdiff_defn}
f(z) \geq f(x) + g^\top (z-x),\quad 
 \forall z \in \mathrm{dom}(f).
\end{equation}
If $x \in \mathrm{Int}(\mathrm{dom}(f))$, 
then subgradient of $f$ at $x$, i.e., $\partial f(x)$ exists.

If a function $f$ is convex and differentiable at $x$, then its gradient at $x$ is a subgradient.  
A function $f$ is called subdifferentiable at $x$ if there exists at least one subgradient at
$x$. 
The set of subgradients of $f$ at the point $x$ is called the subdifferential of $f$ at $x$, and
is denoted $\partial f(x)$. 
A function $f$ is called subdifferentiable if it is subdifferentiable at all $x \in \mathrm{dom}(f)$.

The subdifferential $\partial f(x)$ is always a closed convex set, even if $f$ is not convex. This follows
from the fact that it is the intersection of an infinite set of half-spaces:
\begin{equation*}
\partial f(x) = \bigcap_{z \in \mathrm{dom}(f) } \{ g|f(z) \geq f(x) + g^\top (z-x) \}.
\end{equation*}

A point $x^\star$ is a minimizer of a convex function $f$ if and only if $f$ is subdifferentiable at $x^\star$ and \begin{equation*}
0 \in \partial f(x^\star),
\end{equation*}
that is, $g=0$ is a subgradient of $f$ at $x^\star$. 
This follows directly from the fact that $f(x) \geq f(x^\star)$ for all $x \in \mathrm{dom}(f)$.
\subsubsection{Vector norms and their subdifferentials}
The following are the vector norms on $x \in \R^n$ and the corresponding subdifferentials calculated using the equation \eqref{eq: subdiff_defn}:
\begin{itemize}
	\item \textbf{$\ell_1$ norm:} Let $f(x) = \nmm{x}_1$, then \begin{equation}\label{eq: l1_subdiff}
	[\partial f(x)]_k = \begin{cases}
	\sg(x_k), &\mathrm{if~} x_k \neq 0 \\
	y_k,\text{ where } y_k \in [-1,1], &\mathrm{if~} x_k = 0
	\end{cases}
	\end{equation}
	\item \textbf{$\ell_2$ norm:} Let $f(x) = \nmm{x}_2$, then \begin{equation}\label{eq: l2_subdiff}
	\partial f(x) = \begin{cases}
	\frac{x}{\nmeu{x} }, &\mathrm{if~} x \neq 0 \\
	y, \text{ where } \nmeu{y} \leq 1, &\mathrm{if~} x = 0
	\end{cases}
	\end{equation}
	\item \textbf{$\ell_{\infty}$ norm:} Let $f(x) = \nmm{x}_{\infty}$, then \begin{equation}\label{eq: l_inf_subdiff}
	[\partial f(x) ]_k = \begin{cases}
	\sg(x_k), &\mathrm{if~} x_k = f(x) \\
	0, &\mathrm{else}
	\end{cases}
	\end{equation}
\end{itemize}
where $k = 1,2, \ldots, n $.
\subsection{Karush-Kuhn-Tucker conditions}\label{app: KKT}
The Karush-Kuhn-Tucker (KKT) conditions are first-order necessary conditions 
for a solution to a convex programming to be optimal. Let us consider a convex optimization problem
\begin{equation*}
\begin{aligned}
& \underset{x}{\text{minimize}}
& & f(x)\\
& \text{subject to }
& & g_i(x) \leq 0, \quad i = 1,2, \cdots, m\\
&&& h_j(x) = 0, \quad j = 1,2, \cdots, l
\end{aligned}
\end{equation*}
where $g_i(x)$ are $m$ inequality constraints and $h_j(x)$, are $l$ equality constraints.

Let $x^\star$ is the optimal solution and suppose $f(x)$, $g_i(x)$ and $h_j(x)$, for all $i$ and $j$, be subdifferentiable at $x^\star$. 

The Lagrangian formulation of the problem is given by
\begin{equation*}
\mathcal{L}(x^\star,\beta,\mu) = f(x) + \sum_{i=1}^{m} \mu_i g_i(x) + \sum_{j=1}^{l} \beta_j h_j(x)
\end{equation*}
where $\beta$ and $\mu$ are vectors of multipliers.

The KKT conditions are given as follows:\\
\textbf{Stationarity}
\begin{equation*}
0 \in \partial \mathcal{L}(x^\star,\beta,\mu)
\end{equation*}
\textbf{Primal Feasibility}
\begin{align*}
g_i(x^\star) & \leq 0, \quad \text{for }i = 1,2, \cdots, m\\
h_j(x^\star) & = 0, \quad \text{for }j = 1,2, \cdots, l
\end{align*}
\textbf{Dual Feasibility}
\begin{equation*}
\mu_i \geq 0, \quad \text{for }i = 1,2, \cdots, m
\end{equation*}
\textbf{Complementary slackness}
\begin{equation*}
\mu_i g_i(x^\star) = 0, \quad \text{for }i = 1,2, \cdots, m
\end{equation*}
If the inequality constraints are not active, that is, $g_i(x^\star) < 0$ for some $i$, then the problem is unconstrained with respect to that constraint, that is, 
the corresponding multiplier is zero, or $\mu_i = 0$.
\section{Proof of limiting behavior of CLOT solution} \label{app: proof_conti}
First, let us define the optimization problem \eqref{eq:CLOT_state} in Lagrangian form with $\beta$, $\gamma$ and $\alpha_i$ for all $i = 1, \cdots, N-1$ as the Lagrangian parameters:
\begin{multline}
\label{eq:CLOT_Lag}
L(\hat{u} , \lambda )  \triangleq  h \nmm{\hat{u} }_1 + \lambda \sqrt{h} \nmeu{\hat{u} } \\ 
+\beta \nmeu { A_d^N \xi + \Phi_N \hat{u} } + \gamma (\nmm{\hat{u} }_\infty - 1)  \\
+ \sum_{i=1}^{N-1} \alpha_i (\nmeu{A_d^i \xi + \Psi_N^i \hat{u} }  - \theta ),
\end{multline}
where $\hat{u}$ is the feasible optimal solution and $\Psi_N^i$ is the $i^{th}$ block row of the matrix $\Psi_N$, that is,
\begin{equation}
\Psi_N^i \triangleq \begin{bmatrix} A_d^{i-1}B_d & A_d^{i-2}B_d & \ldots, & B_d & 0 & \ldots & 0\end{bmatrix}
\end{equation}
The KKT conditions for this problem are given by
\begin{equation}
0 \in  \partial f(\hat{u} ) + \beta \partial g(\hat{u} ) + \gamma \partial p(\hat{u} )  + \sum_{i=1}^{N-1} \alpha_i \partial q_i(\hat{u} ),  \label{eq:KKT-1}
\end{equation}
and
\begin{align}
\nmeu { A_d^N \xi + \Phi_N \hat{u} } &  =  0 \label{eq:KKT-2} \\
\nmm{\hat{u} }_\infty & \leq  1  \label{eq:KKT-3} \\
\gamma (\nmm{\hat{u} }_\infty - 1) & = 0 \label{eq:KKT-4} \\
\gamma & \geq 0 \label{eq:KKT-5} \\
\alpha_i (\nmeu{A_d^i \xi + \Psi_N^i \hat{u} }  - \theta ) &  = 0 \label{eq:KKT-6} \\
\alpha_i &  \geq 0 \label{eq:KKT-7}
\end{align}
where $f(\hat{u} ) = h \nmm{\hat{u}}_1 + \lambda \sqrt{h} \nmeu{\hat{u} } $, $g(\hat{u} ) = \nmeu { A_d^N \xi + \Phi_N \hat{u} } $, $p(\hat{u} ) = \nmm{\hat{u} }_\infty$ and $q_i(\hat{u} ) = \nmeu{A_d^i \xi + \Psi_N^i \hat{u} } = \nmeu{\hat{x}_i }$.

Let us consider two components of $\hat{u}$, namely $k$ and $l$, assuming both are not zeros simultaneously. Let us define $a_k:= A_d^{N-k}B_d$,
and $[\Psi_N^i]_k$ as the $k$-th component of $\Psi_N^i$.
Expanding the partial derivatives leads to
\begin{multline}
\beta a_k^\top v_k + h \cdot \sg (\hat{u}_k)
+ \lambda \cdot \sqrt{h} \frac{ \hat{u}_k }{ \nmeu {  \hat{u} } }\\
+ \gamma \cdot \sg (\hat{u}_k) \cdot \delta (|\hat{u}_k| = \nmm {\hat{u}}_\infty)\\
+ \sum_{i=1}^{N-1} \alpha_i [\Psi_N^i]_k^\top \frac{\hat{x}_i }{\nmeu{\hat{x}_i} } = 0
\label{eq:diff_L_k}
\end{multline}
and
\begin{multline}
\beta a_l^\top v_l + h \cdot \sg (\hat{u}_l)
+ \lambda \sqrt{h} \frac{ \hat{u}_l }{ \nmeu {  \hat{u} } }\\
+ \gamma \cdot \sg (\hat{u}_l) \cdot \delta (|\hat{u}_l| = \nmm {\hat{u}}_\infty)\\
+ \sum_{i=1}^{N-1} \alpha_i [\Psi_N^i]_l^\top \frac{\hat{x}_i }{\nmeu{\hat{x}_i} }  = 0
\label{eq:diff_L_l}
\end{multline}
where $v_k$ and $v_l$ are in subdifferential of $\partial \nmeu{.} (A_d^N \xi + \Phi_N \hat{u} )$, that is, since $A_d^N \xi + \Phi_N \hat{u} = 0$, the subgradient is from the set of $\{ v: \nmeu{v} \leq 1 \}$.

From KKT conditions \eqref{eq:KKT-4} and \eqref{eq:KKT-5}, it can be stated that if $\nmm{\hat{u} }_\infty < 1$, then $\gamma = 0$ and if $\nmm{\hat{u} }_\infty = 1$, then $\gamma \geq 0$.

From KKT conditions \eqref{eq:KKT-6} and \eqref{eq:KKT-7}, it can be stated that if $\nmeu{A_d^i \xi + \Psi_N^i \hat{u} } < \theta$, then $\alpha_i = 0$ and if $\nmeu{A_d^i \xi + \Psi_N^i \hat{u} } = \theta$, then $\alpha_i \geq 0$.

Let $l = k + 1$. Then we have
\begin{itemize}
	\item If $\hat{u}_k \neq 0$ and $ \hat{u}_l \neq 0$, then $\sg (\hat{u}_l) = \sg (\hat{u}_k)$.
	\item If $\hat{u}_k = 0 \neq \hat{u}_l $, then $\sg (\hat{u}_k) = 0$
	\item If $\hat{u}_l = 0 \neq \hat{u}_k $, then $\sg (\hat{u}_l) = 0$
\end{itemize} 

And also, $a_k = A_d a_l = (I + M)a_l$, where 
\begin{equation}
M = e^{Ah}-I = \sum_{i=1}^{\infty} \frac{(Ah)^i}{i!}.
\label{eq:M}
\end{equation}
If $k \leq i $, then $[\Psi_N^i]_k = A_d^{i-k}B_d$ and if $k > i$, then  $[\Psi_N^i]_k = 0 \in \mathbb{R}^n$.

Subtracting equations \eqref{eq:diff_L_k} and \eqref{eq:diff_L_l}, when  $\sg(\hat{u}_k) = \sg(\hat{u}_l) $, we have
\begin{multline}\label{eq:k-l_equal}
\beta (a_k^\top v_k - a_l^\top v_l) + \l \sqrt{h} \frac{(\hat{u}_k - \hat{u}_l) }{\nmeu{\hat{u} } } \\
+ \gamma \bigl\{ \sg (u_k) \cdot \delta (|\hat{u}_k| = \nmm {\hat{u}}_\infty) 
- \sg (u_l) \cdot \delta (|\hat{u}_l| = \nmm {\hat{u}}_\infty)\bigr\}\\
 + \sum_{i=1}^{N-1} \alpha_i ([\Psi_N^i]_k - [\Psi_N^i]_l)^\top \frac{\hat{x}_i }{\nmeu{\hat{x}_i } } = 0
\end{multline}
where 
\begin{equation}
[\Psi_N^i]_k - [\Psi_N^i]_{k+1} = 
\begin{cases}
 (A_d - I)[\Psi_N^i]_{k+1}, &\text{ if~ } k \leq i-1,\\
 B_d,  &\text{ if~ } k = i,\\
 0,   &\text{ if~ } k > i.\\
\end{cases}
\end{equation}

Equation \eqref{eq:k-l_equal} when  $\sg(\hat{u}_k) \neq \sg(\hat{u}_l) $, i.e., $\hat{u}_l = 0 \neq \hat{u}_k $ becomes
\begin{multline}\label{eq:k-l_unequal}
\beta (a_k^\top v_k - a_l^\top v_l) + h \cdot \sg (\hat{u}_k) + \l \sqrt{h} \frac{\hat{u}_k }{\nmeu{\hat{u} } }\\
+ \gamma \cdot  \sg (u_k) \cdot \delta (|\hat{u}_k| = \nmm {\hat{u}}_\infty)  \\ 
 + \sum_{i=1}^{N-1}  \alpha_i ([\Psi_N^i]_k - [\Psi_N^i]_l)^\top \frac{\hat{x}_i }{\nmeu{\hat{x}_i } } = 0
\end{multline}
The other case is similar to this one.

We also need to consider the following cases with respect to $\alpha_i$:
\begin{enumerate}
	\item[a.] $\nmeu{\hat{x}_i } < \theta$, $\alpha_i = 0$.
	\item[b.] $\nmeu{\hat{x}_i } = \theta$, $\alpha_i > 0$.
\end{enumerate}	
We need to consider the following cases for \eqref{eq:k-l_equal} with respect to $\gamma$:
\begin{enumerate}
	\item $\nmm{\hat{u} }_\infty = 1$, $|\hat{u}_k| < 1 $ and $|\hat{u}_l| < 1 $. 
	In this case, the terms with $\gamma$ both become zero.
	\item $\nmm{\hat{u} }_\infty = 1$, $|\hat{u}_k| = 1 $ and $|\hat{u}_l| = 1 $. 
	In this case, the terms with $\gamma$ will become $\sg (\hat{u}_k ) - \sg (\hat{u}_l )$ and since 
	$\sg (\hat{u}_k ) = \sg (\hat{u}_l )$,
	 the term $\gamma (\sg (\hat{u}_k ) - \sg (\hat{u}_l ))$ vanishes.
	\item $\nmm{\hat{u} }_\infty = 1$; $|\hat{u}_k| = 1 $ and $|\hat{u}_l| < 1 $, or $|\hat{u}_k| < 1 $ and $|\hat{u}_l| = 1 $. In this case, only one term of $\gamma$ remains, and the other goes to zero. Thus, it is either $\sg (\hat{u}_k )$ or $\sg (\hat{u}_l )$ respectively.	
\end{enumerate}
Thus, in cases 1 and 2, the equation \eqref{eq:k-l_equal} becomes
\begin{multline}\label{eq:cases_1_2}
\beta (a_k^\top v_k - a_l^\top v_l) + \l \sqrt{h} \frac{(\hat{u}_k - \hat{u}_l) }{\nmeu{\hat{u} } }\\ + \sum_{i=1}^{N-1} \alpha_i ([\Psi_N^i]_k - [\Psi_N^i]_l)^\top \frac{\hat{x}_i }{\theta }  = 0
\end{multline}
for $i$ which satisfy the case `b', that is $\nmeu{\hat{x}_i} = \theta$.
But in case 3, the equation \eqref{eq:k-l_equal} becomes either of the following:
\begin{multline}
\beta (a_k^\top v_k - a_l^\top v_l) + \l \sqrt{h} \frac{(\hat{u}_k - \hat{u}_l) }{\nmeu{\hat{u} } } + \gamma \cdot \sg (\hat{u}_k )\\
+ \sum_{i=1}^{N-1} \alpha_i ([\Psi_N^i]_k - [\Psi_N^i]_l)^\top \frac{\hat{x}_i }{\theta } = 0 \label{eq:cases_3_k}
\end{multline}
or
\begin{multline}
\beta (a_k^\top v_k - a_l^\top v_l) + \l \sqrt{h} \frac{(\hat{u}_k - \hat{u}_l) }{\nmeu{\hat{u} } } - \gamma \cdot \sg (\hat{u}_l )\\
+ \sum_{i=1}^{N-1} \alpha_i ([\Psi_N^i]_k - [\Psi_N^i]_l)^\top \frac{\hat{x}_i }{\theta }  = 0 \label{eq:cases_3_l}
\end{multline}
for $i$ which satisfy the case `b', that is $\nmeu{\hat{x}_i} = \theta$.
We need to consider the following cases for \eqref{eq:k-l_unequal} -
\begin{enumerate}
	\item $\nmm{\hat{u} }_\infty = 1$ and $|\hat{u}_k| < 1 $. In this case, the term of $\gamma(\sg (\hat{u}_k ) - \sg (\hat{u}_l ) )$ vanishes. Therefore, this case is similar to case 1 for \eqref{eq:k-l_equal}.\\
	\item $\nmm{\hat{u} }_\infty = 1$ and $|\hat{u}_k| = 1 $. In this case, the term of $\gamma$ becomes $\gamma \sg(\hat{u}_k )$. Therefore, this case is similar to case 3 for \eqref{eq:k-l_equal}.
\end{enumerate}
Thus, in case 1, the equation \eqref{eq:k-l_unequal} becomes
\begin{multline}\label{eq:case_1_ue}
\beta (a_k^\top v_k - a_l^\top v_l) + h \cdot\sg(\hat{u}_k ) + \l \sqrt{h} \frac{\hat{u}_k }{\nmeu{\hat{u} } }\\
+ \sum_{i=1}^{N-1} \alpha_i ([\Psi_N^i]_k - [\Psi_N^i]_l)^\top \frac{\hat{x}_i }{\theta }  = 0
\end{multline}
for $i$ which satisfy the case `b', that is $\nmeu{\hat{x}_i} = \theta$.
In case 2, the equation \eqref{eq:k-l_unequal} becomes
\begin{multline}\label{eq:case_2_ue}
\beta (a_k^\top v_k - a_l^\top v_l) + (h+ \gamma) \cdot \sg(\hat{u}_k ) + \l \sqrt{h} \frac{\sg(\hat{u}_k)  }{\nmeu{\hat{u} } }\\
+ \sum_{i=1}^{N-1}\alpha_i ([\Psi_N^i]_k - [\Psi_N^i]_l)^\top \frac{\hat{x}_i }{\theta }  = 0
\end{multline}
for $i$ which satisfy the case `b', that is $\nmeu{\hat{x}_i} = \theta$.
On observation, we can say that \eqref{eq:cases_3_k}, \eqref{eq:cases_3_l}, \eqref{eq:case_1_ue} and \eqref{eq:case_2_ue} are similar.

Let us move further with case 3 with equation \eqref{eq:cases_3_k} 
(the other cases can be derived from this case):
\begin{multline}
\l \sqrt{h} \frac{(\hat{u}_k - \hat{u}_l) }{\nmeu{\hat{u} } }  = -\beta (a_k^\top v_k - a_l^\top v_l) - \gamma (\sg (\hat{u}_k ) )\\
- \sum_{i=1}^{N-1} \alpha_i ([\Psi_N^i]_k - [\Psi_N^i]_l)^\top \frac{\hat{x}_i }{\theta }
\end{multline}
From this, we have
\begin{multline}\label{eq:before_cases_psi}
\l \sqrt{h} \frac{|\hat{u}_k - \hat{u}_l| }{\nmeu{\hat{u} } }  \leq |\beta (a_k^\top v_k - a_l^\top v_l)| +  \gamma\\
+ \sum_{i=1}^{N-1} \frac{\alpha_i}{\theta} |([\Psi_N^i]_k - [\Psi_N^i]_l)^\top \hat{x}_i |
\end{multline}
Therefore depending upon $i$ and $k$, \eqref{eq:before_cases_psi} can become either of the following:
\begin{multline}\label{eq:cases_psi_1}
\l \sqrt{h} \frac{|\hat{u}_k - \hat{u}_l| }{\nmeu{\hat{u} } }  \leq |\beta (a_k^\top v_k - a_l^\top v_l)| +  \gamma\\
+ \sum_{i=1}^{N-1} \frac{\alpha_i}{\theta } |((A_d - I) [\Psi_N^i]_{k+1} )^\top \hat{x}_i | 
\end{multline}
\begin{equation}
\l \sqrt{h} \frac{|\hat{u}_k - \hat{u}_l| }{\nmeu{\hat{u} } }  \leq |\beta (a_k^\top v_k - a_l^\top v_l)| + \gamma
+ \sum_{i=1}^{N-1} \frac{\alpha_i}{\theta } |B_d ^\top \hat{x}_i | \label{eq:cases_psi_2}
\end{equation}
or
\begin{equation}
\l \sqrt{h} \frac{|\hat{u}_k - \hat{u}_l| }{\nmeu{\hat{u} } }  \leq |\beta (a_k^\top v_k - a_l^\top v_l)| +  \gamma 
\label{eq:cases_psi_3}
\end{equation}

Let us move forward with \eqref{eq:cases_psi_1}, as the others can be derived from this one.
\begin{equation}
\begin{split}
\l \sqrt{h} \frac{|\hat{u}_k - \hat{u}_l| }{\nmeu{\hat{u} } }  & \leq  |\beta (a_k^\top v_k - a_l^\top v_l)| +  \gamma\\
&\quad + \sum_{i=1}^{N-1} \frac{\alpha_i}{\theta } \nmeu{(A_d - I) [\Psi_N^i]_{k+1} } \nmeu{\hat{x}_i}\\
& \quad \qquad\text{(by Cauchy-Schwarz inequality) }\\
&\leq  |\beta ((a_l + M a_l)^\top v_k - a_l^\top v_l)| +  \gamma\\
&\quad + \sum_{i=1}^{N-1} \frac{\alpha_i}{\theta} \nmeu{(A_d - I) [\Psi_N^i]_{k+1} } \nmeu{\hat{x}_i}\\
&\leq  |\beta ( (M a_l)^\top v_k + a_l^\top (v_k - v_l)| +  \gamma\\
&\quad + \sum_{i=1}^{N-1} \alpha_i \nmm{(A_d - I)} \cdot \nmeu{[\Psi_N^i]_{k+1} } \\
&\leq  |\beta| ( \nmm{M} \nmeu{a_l} \nmeu{v_k} + \nmeu{a_l} \nmeu{v_k - v_l} )\\ 
&\quad +  \gamma + \sum_{i=1}^{N-1} \alpha_i \nmm{M} \cdot \nmeu{A_d^{i-k-1}B_d } \\
&\leq  |\beta| \cdot (\nmm{M}+2) \cdot \nmeu{a_l} +  \gamma \\
&\quad + \sum_{i=1}^{N-1} \alpha_i \nmm{M} \cdot \nmeu{A_d^{i-k-1}B_d }
\end{split}
\end{equation}
This leads to
\begin{equation}\label{eq:abs_diff_req}
\begin{split}
|\hat{u}_k - \hat{u}_l|  
 &\leq  \frac{|\beta| \cdot (\nmm{M} + 2) \cdot \nmeu{a_l} \cdot \nmeu{\hat{u} } }{\l \sqrt{h} }\\
 &\quad + \frac{\gamma \cdot \nmeu{\hat{u} } }{\l \sqrt{h} } \\
 &\quad + \frac{\nmm{M} \cdot \nmeu{ \hat{u} } }{\lambda \sqrt{h} } \sum_{i=1}^{N-1} \alpha_i \cdot \nmeu{A_d^{i-k-1}B_d } 
\end{split}
\end{equation}
To obtain the order of $|\hat{u}_k - \hat{u}_l|$, 
we need to know the order of the RHS of equation \eqref{eq:abs_diff_req} in terms of $h$. 
For this, we consider $\nmeu{\hat{u} }$, $|\beta|$, $\gamma$, $\nmm{M}$, $\nmeu{a_l}$, $\alpha_i$ and $\nmeu{A_d^{i-k-1}B_d }$.

First, from \eqref{eq:M}, $\nmm{M}$ is $O(h)$.
Next, we have
\begin{align*}
a_l & =  A_d^{N-k-1} B_d \\
& =  e^{Ah(N-k-1) } \cdot \sum_{i =  0}^{\infty} \frac{A^i h^{i+1} }{(i+1)! } \cdot B \\
& =  e^{AT} \cdot e^{-A(k+1)h } \cdot \sum_{i =  0}^{\infty} \frac{A^i h^{i+1} }{(i+1)! } \cdot B,
\end{align*}
thus $\nmeu{a_l}$ is of $O(h)$.
Similarly, $\nmeu{A_d^{i-k-1}B_d }$ is of $O(h)$. 
From equation \eqref{eq:KKT-2},
we have
\[
A_d^N \xi + \Phi_N \hat{u} = 0,
\]
and hence
\begin{equation}
\nmeu{A_d^N \xi} = \nmeu{\Phi_N \hat{u}} 
 \leq \nmm{\Phi_N} \cdot \nmeu{\hat{u} }.
\end{equation}
Since $(A_d,B_d)$ is controllable, we have $\nmm{\Phi_N}>0$,
which gives
\begin{equation}\label{eq:B.3.2}
\frac{\nmeu{A_d^N \xi} }{\nmm{\Phi_N} }\leq \nmeu{\hat{u} }
\end{equation}
Also, from equation \eqref{eq:KKT-3}, we have
\begin{equation}\label{eq:B.4.2}
\nmeu{\hat{u} } \leq \sqrt{N} \nmm{\hat{u} }_\infty \leq \sqrt{N} = \sqrt{ \frac{T}{h} }
\end{equation}
It follows from \eqref{eq:B.3.2} and \eqref{eq:B.4.2} that
\begin{equation} \label{eq:u2_bounds}
\frac{\nmeu{A_d^N \xi} }{\nmm{\Phi_N} } \leq \nmeu{\hat{u} } \leq \sqrt{ \frac{T}{h} }.
\end{equation}
Therefore, we can conclude that $\nmeu{\hat{u} }$ is $O(\frac{1}{\sqrt{h} } )$.
Also, from \eqref{eq:AdBd} and \eqref{eq:Phi_N}, 
$\nmm{\Phi_N}$ is of $O(h)$.

Then, let us consider the equations \eqref{eq:diff_L_k} and \eqref{eq:diff_L_l} in case 2, that is, $\nmm{\hat{u} }_\infty = |\hat{u}_k| = |\hat{u}_l| = 1$.
From \eqref{eq:diff_L_k}, we have
\begin{multline}\label{eq:beta_alpha_k}
\beta = -\sg (\hat{u}_k) \Bigg(\frac{h \nmeu{\hat{u} } + \lambda \sqrt{h} }{\nmeu{\hat{u} } \cdot (a_k^\top v_k) } \Bigg) \\
 - \sum_{i=1}^{N-1} \frac{\alpha_i}{\theta (a_k^\top v_k) } [\Psi_N^i]_k^\top \hat{x}_i
\end{multline}
and from \eqref{eq:diff_L_l}, we have
\begin{multline}\label{eq:beta_alpha_l}
\beta = -\sg (\hat{u}_l) \Bigg(\frac{h \nmeu{\hat{u} } + \lambda \sqrt{h} }{\nmeu{\hat{u} } \cdot (a_l^\top v_l) } \Bigg)\\
 - \sum_{i=1}^{N-1} \frac{\alpha_i}{\theta (a_l^\top v_l) } [\Psi_N^i]_l^\top \hat{x}_i.
\end{multline}
By equating \eqref{eq:beta_alpha_k} and \eqref{eq:beta_alpha_l}, we have
\begin{multline*}
\sum_{i=1}^{N-1} \frac{\alpha_i}{\theta } \Biggl(\frac{[\Psi_N^i]_k}{a_k^\top v_k } - \frac{[\Psi_N^i]_l}{a_l^\top v_l } \Biggl)^\top \hat{x}_i\\ 
= \sg (\hat{u}_l) \Bigg(\frac{h \nmeu{\hat{u} } + \lambda \sqrt{h} }{\nmeu{\hat{u} } } \Bigg) \cdot \Bigg( \frac{1}{a_l^\top v_l} - \frac{1}{a_k^\top v_k} \Bigg),
\end{multline*}
from which we have
\begin{multline*}
\sum_{i=1}^{N-1} \frac{\alpha_i}{\theta } |( a_l^\top v_l [\Psi_N^i]_k - a_k^\top v_k [\Psi_N^i]_l)^\top \hat{x}_i|  \leq\\ 
 \Bigg(\frac{h \nmeu{\hat{u} } + \lambda \sqrt{h} }{\nmeu{\hat{u} } } \Bigg) |a_k^\top v_k - a_l^\top v_l|
\end{multline*}
Without loss of generality, let us assume there is only one $i \in \{1,2,\ldots,N-1\}$
which satisfies case `b', that is,
\begin{equation}\label{eq:alpha_i}
\begin{split}
\alpha_i 
& \leq \frac{\theta (h \nmeu{\hat{u} } + \lambda \sqrt{h}) }{\nmeu{\hat{u} } }
\frac{|a_k^\top v_k - a_l^\top v_l|}{|( a_l^\top v_l [\Psi_N^i]_k - a_k^\top v_k [\Psi_N^i]_l)^\top \hat{x}_i|} \\
&\leq \frac{\theta (h \sqrt{\frac{T}{h} } + \lambda \sqrt{h}) }{\frac{\nmeu{A_d^N \xi} }{\nmm{\Phi_N} } }
\frac{(2+\nmm{M} ) \cdot \nmeu{a_l} }{|( a_l^\top v_l [\Psi_N^i]_k - a_k^\top v_k [\Psi_N^i]_l)^\top \hat{x}_i|}
\end{split}
\end{equation}
If $i\geq k+1$, we have
\begin{align*}
&|( a_l^\top v_l [\Psi_N^i]_k - a_k^\top v_k [\Psi_N^i]_l)^\top \hat{x}_i|\\
&~~\leq \nmeu{a_l^\top v_l [\Psi_N^i]_k - a_k^\top v_k [\Psi_N^i]_l}\nmeu{\hat{x}_i}\\
&~~\leq \theta  \nmeu{a_l^\top v_l A_d - (a_l^\top v_k + (M a_l)^\top v_k)I } \cdot \nmeu{A_d^{i-k-1}B_d } \\
&~~\leq 2\theta (1+\nmm{M} ) \nmeu{a_l} \cdot \nmeu{A_d^{i-k-1}B_d }
\end{align*}
Therefore, $|( a_l^\top v_l [\Psi_N^i]_k - a_k^\top v_k [\Psi_N^i]_l)^\top \hat{x}_i|$ is of $O(h^2)$.
By plugging this result into \eqref{eq:alpha_i}, we get $\alpha_i$ is of $O(\sqrt{h} )$.
Thus, from \eqref{eq:beta_alpha_k}, we have
\begin{equation}\label{eq:beta}
|\beta| \leq \frac{\sqrt{h} (\l + \sqrt{T} ) \cdot \nmeu{\Phi_N} }{\nmeu{A_d^N \xi} \cdot |a_l^\top v_l| } + \sum_{i=1}^{N-1} \frac{\alpha_i \nmeu{A_d^{i-k-1} B_d} }{|a_l^\top v_l|}
\end{equation}
Therefore, $\beta$ is of $O(\sqrt{h} )$.

By considering equations \eqref{eq:diff_L_k} and \eqref{eq:diff_L_l} in case 3, 
we have respectively
\begin{multline}
\beta = \frac{- \sg (\hat{u}_k) }{a_k^\top v_k \cdot \nmeu{\hat{u} } } \bigl( (h+\gamma) \nmeu{\hat{u} } + \lambda \sqrt{h} \bigr) \\
- \frac{1}{a_k^\top v_k}\sum_{i \in [N-1]} \frac{\alpha_i}{\theta } \cdot [\Psi_N^i]_k^\top \hat{x}_i \label{eq:beta_k}
\end{multline}
and
\begin{multline}
\beta = \frac{- \sg (\hat{u}_l ) }{a_l^\top v_l \cdot \nmeu{\hat{u} } } (h \nmeu{\hat{u} } + \l \sqrt{h} |\hat{u}_l| ) \\
- \frac{1}{a_l^\top v_l}\sum_{i \in [N-1]} \frac{\alpha_i}{\theta } \cdot [\Psi_N^i]_l^\top \hat{x}_i. 
\label{eq:beta_l}
\end{multline}
Equating $\beta$ from both the equations, we get
\begin{align}
\gamma 
&\leq \frac{(h \nmeu{\hat{u} } + \l \sqrt{h} )}{\nmeu{\hat{u} } } \left|\frac{a_k^\top v_k - a_l^\top v_l}{a_l^\top v_l}\right|\\
&\quad + \sum_{i=1}^{N-1} \frac{\alpha_i}{\theta } \cdot \left|\Big( \frac{a_k^\top v_k [\Psi_N^i]_l  - a_l^\top v_l [\Psi_N^i]_k}{ a_l^\top v_l } \Big)^\top \hat{x}_i\right|\nonumber \\
&\leq  (h \sqrt{\frac{T}{h} } + \l \sqrt{h} ) \frac{(2+\nmm{M}) \cdot \nmeu{a_l} }{|a_l^\top v_l| \cdot \nmeu{\hat{u} } }  \nonumber \\
&\quad + \sum_{i=1}^{N-1} \frac{2\alpha_i(1+\nmm{M} ) \nmeu{a_l} \cdot \nmeu{A_d^{i-k-1}B_d }}{ |a_l^\top v_l|}
  \nonumber \\
&\leq  \sqrt{h}(\sqrt{T} + \l) \frac{ (2+\nmm{M}) \cdot \nmeu{a_l} \cdot \nmm{\Phi_N} }{|a_l^\top v_l| \cdot \nmeu{A_d^N \xi}   } \nonumber \\
&\quad + \sum_{i=1}^{N-1} \frac{ 2\alpha_i(1+\nmm{M} ) \nmeu{a_l} \cdot \nmeu{A_d^{i-k-1}B_d }}{ |a_l^\top v_l|} \label{eq:gamma} 
\end{align}
Since, $\nmm{M}$ is of $O(h)$, $\nmeu{a_l}$, $\nmeu{A_d^{i-k-1}B_d }$,  $|a_l^\top v_l|$ and $ \nmm{\Phi_N }$ are of $O(h)$ and $\alpha_i$ is of $O(\sqrt{h} )$, we get $\gamma$ is of $O(h \sqrt{h} )$.
Thus, \eqref{eq:abs_diff_req} becomes, 
\begin{multline*}
|\hat{u}_k - \hat{u}_l| \leq  \frac{|\beta| \sqrt{T} \cdot (\nmm{M} + 2) \cdot \nmeu{a_l}  }{\l h } + \frac{\gamma \sqrt{T} }{\l h }\\
 + \frac{\nmm{M} \cdot \sqrt{T} }{\l h } \sum_{i=1}^{N-1} \alpha_i \cdot \nmeu{A_d^{i-k} B_d }
\end{multline*}
and by considering all the orders, we get $|\hat{u}_k - \hat{u}_l|$ is of $O(\sqrt{h} )$. Thus as $h \rightarrow 0$, $|\hat{u}_k - \hat{u}_l| \rightarrow 0$.


\bibliographystyle{IEEEtran}

\bibliography{IEEEabrv,reference}
\end{document}